\documentclass[11pt, logo, nonumbering]{preprint}
\usepackage[utf8]{inputenc}
\usepackage{microtype}
\usepackage{graphicx}
\usepackage{amsmath}
\usepackage{booktabs}
\usepackage{xcolor}
\definecolor{mydarkblue}{rgb}{0,0.08,0.45}
\usepackage[colorlinks=true,linkcolor=mydarkblue,citecolor=mydarkblue,filecolor=mydarkblue,urlcolor=mydarkblue]{hyperref}
\usepackage{xspace}
\usepackage{cleveref}
\usepackage{float}
\usepackage{adjustbox}
\usepackage{subcaption}
\usepackage{fancyvrb}

\usepackage{pifont}

\usepackage{tcolorbox}
\newtcolorbox{codebox}{
    colback=green!10,
    colframe=green!50!black,
    boxrule=2pt,
    arc=5pt,
    left=10pt,
    right=10pt,
    top=10pt,
    bottom=10pt,
    width=0.85\textwidth,
    center
}

\usepackage[style=numeric-comp,sorting=none,maxbibnames=3,minnames=1,backend=bibtex, doi=false,eprint=false,url=false]{biblatex}

\definecolor{gred}{RGB}{250, 210, 207}
\definecolor{coolblue1}{rgb}{0.91, 0.94, 0.98}
\definecolor{coolblue2}{rgb}{0.76, 0.85, 0.94}
\definecolor{coolblue3}{rgb}{0.54, 0.72, 0.87}
\definecolor{coolblue4}{rgb}{1, 1, 1}

\begin{document}

\title{LoCoBench-Agent: An Interactive Benchmark for LLM Agents in Long-Context Software Engineering} 

\author{
Jielin Qiu, Zuxin Liu, Zhiwei Liu, Rithesh Murthy, Jianguo Zhang, Haolin Chen, 
\\ 
Shiyu Wang, Ming Zhu, Liangwei Yang, Juntao Tan, Roshan Ram, Akshara Prabhakar, \\
Tulika Awalgaonkar, Zixiang Chen, Zhepeng Cen, Cheng Qian, 
\\
Shelby Heinecke, Weiran Yao, Silvio Savarese, Caiming Xiong, Huan Wang
\\
\textbf{Salesforce AI Research} 
\\
\href{https://github.com/SalesforceAIResearch/LoCoBench-Agent}{
  \raisebox{-0.3\height}{\includegraphics[height=0.8cm]{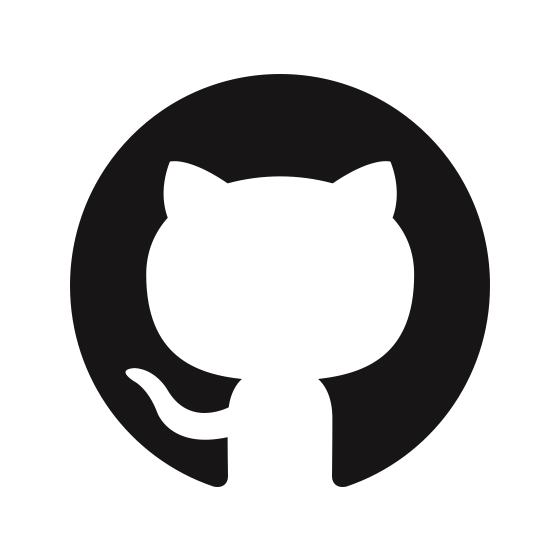}}
  \textbf{SalesforceAIResearch/LoCoBench-Agent}
}
}

\maketitle

\begin{figure}[H]
  \centering
  \includegraphics[width=0.99\textwidth]{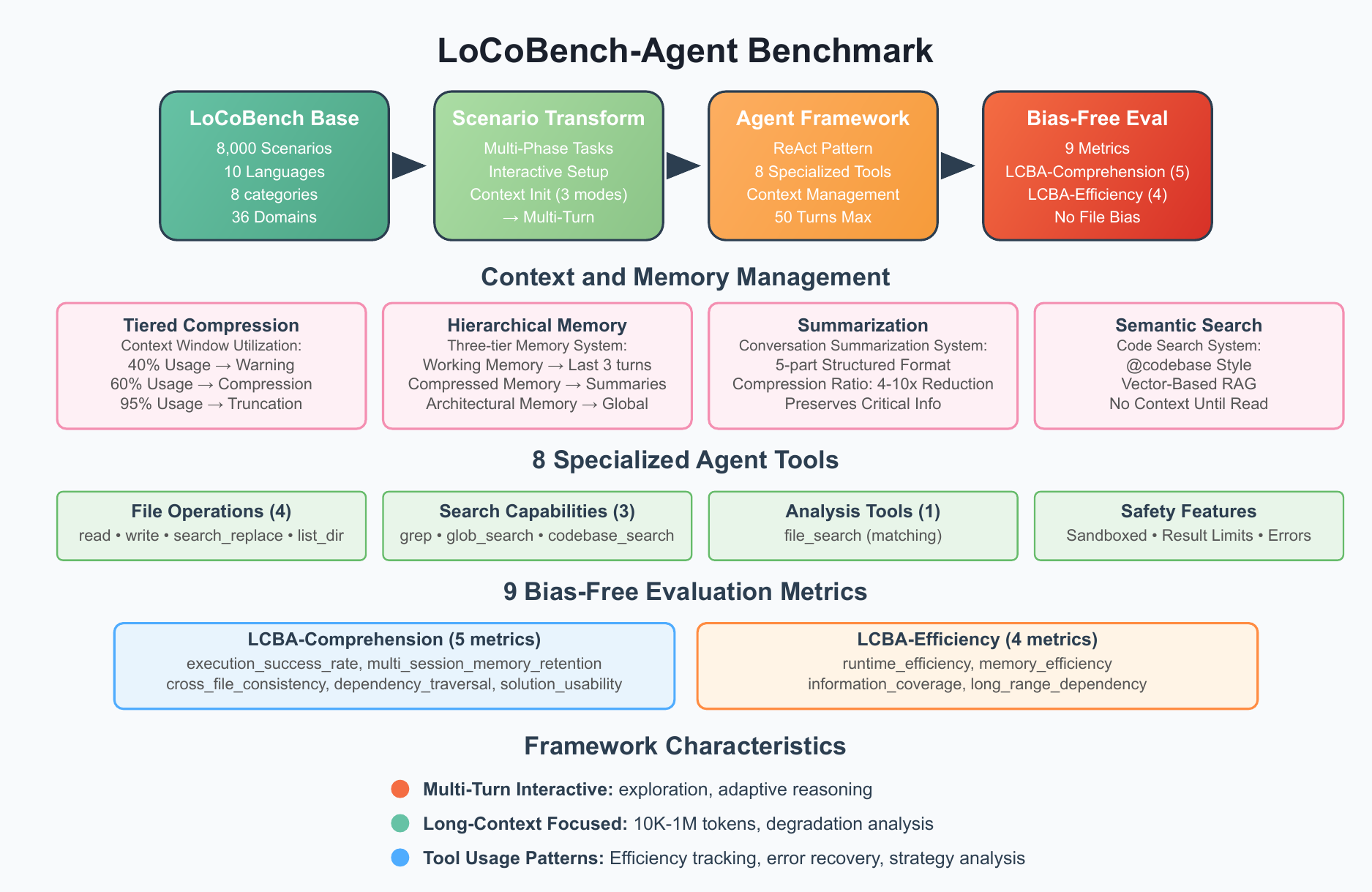}
  \label{fig:locobench_agent_pipeline}
\end{figure}

\begin{abstract}
As large language models (LLMs) evolve into sophisticated autonomous agents capable of complex software development tasks, evaluating their real-world capabilities becomes critical. While existing benchmarks like LoCoBench~\cite{qiu2025locobench} assess long-context code understanding, they focus on single-turn evaluation and cannot capture the multi-turn interactive nature, tool usage patterns, and adaptive reasoning required by real-world coding agents. We introduce \textbf{LoCoBench-Agent}, a comprehensive evaluation framework specifically designed to assess LLM agents in realistic, long-context software engineering workflows. Our framework extends LoCoBench's 8,000 scenarios into interactive agent environments, enabling systematic evaluation of multi-turn conversations, tool usage efficiency, error recovery, and architectural consistency across extended development sessions. We also introduce an evaluation methodology with 9 metrics across comprehension and efficiency dimensions. Our framework provides agents with 8 specialized tools (file operations, search, code analysis) and evaluates them across context lengths ranging from 10K to 1M tokens, enabling precise assessment of long-context performance. Through systematic evaluation of state-of-the-art models, we reveal several key findings: (1) agents exhibit remarkable long-context robustness; (2) comprehension-efficiency trade-off exists with negative correlation, where thorough exploration increases comprehension but reduces efficiency; and (3) conversation efficiency varies dramatically across models, with strategic tool usage patterns differentiating high-performing agents. As the first long-context LLM agent benchmark for software engineering, LoCoBench-Agent establishes a rigorous foundation for measuring agent capabilities, identifying performance gaps, and advancing autonomous software development at scale.
\end{abstract}

\section{Introduction}
\label{sec:intro}

The evolution of large language models (LLMs) from passive code completion tools to autonomous software engineering agents represents a fundamental shift in AI-assisted development. Modern LLM agents can now engage in multi-turn conversations, utilize diverse toolsets, adapt strategies based on feedback, and maintain context across extended development sessions~\cite{qian2024agentcoder,hong2023metagpt,yang2024swe}. However, evaluating these sophisticated capabilities requires fundamentally different approaches than traditional single-turn code generation benchmarks.

\textbf{The Agent Evaluation Gap.} While LoCoBench~\cite{qiu2025locobench} established comprehensive evaluation for long-context code understanding, it focuses on single-turn evaluation where models receive complete context and generate responses in isolation. Real-world software development agents operate fundamentally differently: they engage in multi-turn dialogues, incrementally gather information through tool usage, adaptively refine solutions based on feedback, and maintain architectural consistency across extended sessions. Existing agent benchmarks like AgentBench~\cite{liu2023agentbench} and SWE-Bench~\cite{jimenez2023swe} evaluate limited aspects of agent behavior but lack systematic assessment of long-context capabilities, comprehensive tool usage patterns, and bias-free evaluation methodologies.

\textbf{Critical Challenges in Agent Evaluation.} Evaluating LLM agents in software engineering presents unique challenges: (1) \textit{Multi-Turn Complexity:} Agents must maintain context, architectural understanding, and solution coherence across dozens of conversation turns; (2) \textit{Tool Usage Patterns:} Effective agents efficiently utilize file operations, search tools, and code analysis capabilities rather than relying on raw context memory; (3) \textit{Evaluation Bias:} Traditional metrics often exhibit file count bias (rewarding solutions that modify more files); (4) \textit{Long-Context Degradation:} Agent performance may degrade as context length increases, but existing benchmarks cannot systematically measure this effect.

\textbf{Our Contribution: LoCoBench-Agent.} We introduce LoCoBench-Agent, a comprehensive evaluation framework designed for LLM agents in long-context software engineering. Building upon LoCoBench's 8,000 high-quality scenarios, we transform them into interactive agent environments that enable realistic multi-turn evaluation. Our framework introduces:

\begin{itemize}
\item \textbf{Interactive Agent Environment:} We provide a complete agent framework with 8 specialized tools (file operations, search, code analysis) that enable realistic software development workflows across 10K-1M token contexts.

\item \textbf{Bias-Free Evaluation Methodology:} Through iterative metric design and rigorous validation, we developed 9 evaluation metrics (5 comprehension + 4 efficiency) that eliminate file count bias.

\item \textbf{Comprehensive Multi-Turn Assessment:} Our framework evaluates conversation efficiency, tool usage patterns, error recovery speed, cross-file consistency, and long-range dependency resolution, which are the capabilities essential for real-world agents but unmeasured by existing benchmarks.

\item \textbf{Systematic Long-Context Evaluation:} We provide a scalable framework for evaluating state-of-the-art models across all 8,000 scenarios, enabling systematic analysis of how agent performance varies with context length and task complexity.
\end{itemize}

\textbf{Key Findings.} Through comprehensive evaluation, we find insights into agent capabilities and architectural trade-offs: (1) \textit{Comprehension-Efficiency Trade-off:} A fundamental architectural tension exists ($r = -0.42$ negative correlation) where thorough codebase exploration necessary for high comprehension directly conflicts with efficiency optimization, where no current architecture resolves this trade-off, with best-performing agents clustering along a Pareto frontier rather than achieving simultaneous optimization; (2) \textit{Strategic Differentiation:} Conversation efficiency varies dramatically across models (10-22 turns average), with high-performing agents employing distinct strategies, where semantic search first with targeted reads versus exhaustive exploration patterns, revealing that tool usage patterns, not raw capabilities, differentiate agent effectiveness in long-context scenarios.

\section{Design of LoCoBench-Agent}
\label{sec:method}

LoCoBench-Agent transforms LoCoBench's 8,000 static evaluation scenarios into interactive agent environments. The benchmark design consists of four components: (1) scenario transformation from static to interactive format, (2) an intelligent context management system for long-context handling, (3) an interactive agent system with specialized development tools, and (4) a comprehensive multi-turn evaluation pipeline.

\subsection{From LoCoBench to LoCoBench-Agent}

\textbf{Systematic Scenario Conversion Pipeline.} We transform LoCoBench's 8,000 single-turn scenarios into multi-turn interactive environments through a structured three-stage process with caching for efficiency.

\textit{Stage 1: Project Extraction and Normalization.} For each scenario, we extract the complete project codebase from LoCoBench's generation pipeline output, preserving the original file structure, dependency relationships, and architectural organization. Projects span 10 programming languages (Python, JavaScript, Java, C++, Go, Rust, TypeScript, PHP, Ruby, C\#) across 36 domain categories (web applications, machine learning systems, data processing pipelines, system utilities, etc.). Each project contains 10-100 files with 100-10,000 lines of code, generating contexts ranging from 10K to 1M tokens based on difficulty level. We normalize file paths, validate compilation/execution, and create structured metadata including file dependency graphs, entry points, and architectural patterns.

\textit{Stage 2: Task Decomposition and Phase Generation.} Single-turn task specifications transform into multi-phase conversation structures. Complex requirements decompose into incremental sub-tasks requiring exploration ("Understand the authentication system architecture"), planning ("Design the new OAuth2 integration approach"), implementation ("Implement token refresh mechanism"), and validation ("Test the integration with existing endpoints"). Each phase defines: initial prompts, expected agent actions (file reads, tool usage patterns), success conditions (intermediate goals), and dynamic follow-up prompts based on agent progress. This decomposition mirrors real developer workflows where tasks unfold incrementally through discovery and iterative refinement.

\textit{Stage 3: Success Criteria and Evaluation Schema.} We generate multi-faceted success criteria evaluating: (1) \textit{Functional Correctness:} Code compiles, executes, and produces expected outputs; (2) \textit{Architectural Consistency:} Maintains existing design patterns, follows project conventions, preserves module boundaries; (3) \textit{Test Passage:} Existing test suites remain passing, new tests cover added functionality; (4) \textit{Requirement Satisfaction:} Implements all specified features, handles edge cases, includes documentation. Additionally, we define intermediate checkpoints assessing tool usage efficiency, code exploration patterns, and problem-solving approaches throughout the session.

\textbf{Category-Specific Adaptations.} LoCoBench's 8 task categories undergo targeted transformations: \textit{Code Comprehension} $\rightarrow$ \textit{Interactive Code Exploration} where agents incrementally discover codebase structure through targeted file reads and searches rather than receiving all code upfront; \textit{Architectural Understanding} $\rightarrow$ \textit{Interactive Architecture Exploration} requiring multi-turn analysis of design patterns, dependency relationships, and component interactions; \textit{Bug Investigation} $\rightarrow$ \textit{Interactive Debugging Sessions} with hypothesis generation, testing, and iterative refinement; \textit{Feature Implementation} $\rightarrow$ \textit{Collaborative Feature Development} with explicit planning, implementation, and validation phases; \textit{Cross-File Refactoring} $\rightarrow$ \textit{Guided Multi-File Refactoring} requiring coordination across multiple files while maintaining consistency; \textit{Integration Testing} $\rightarrow$ \textit{Test-Driven Development Sessions} where agents write tests before implementation; \textit{Security Analysis} $\rightarrow$ \textit{Interactive Security Auditing} with vulnerability discovery and remediation; \textit{Multi-Session Development} $\rightarrow$ \textit{Extended Development Projects} simulating work across multiple days with context retention.

\textbf{Context Initialization Strategies.} We provide three initialization modes reflecting different real-world scenarios: (1) \textit{Minimal Mode} (default, 90\% of evaluations): Loads only README, project file structure (file names and paths without content), and entry point file names. Agents must discover relevant code through \texttt{file\_search}, \texttt{grep}, \texttt{codebase\_search}, and \texttt{read\_file} tools, mirroring how developers approach unfamiliar codebases in production environments like VS Code or Cursor. This mode prevents context window overflow, encourages efficient tool usage, and provides the most realistic evaluation. (2) \textit{Empty Mode} (exploratory evaluation): Provides only task specification and project root path. Agents must discover everything including file structure, forcing complete exploration. Tests agent capability to work with zero initial context. (3) \textit{Full Mode} (baseline comparison): Loads complete codebase into initial context (all files), simulating traditional single-turn evaluation. Only viable for small projects (<100K tokens) due to context limits. Used for comparing interactive vs. single-turn approaches.

\subsection{Long-Context Memory Management}

Our memory management system draws inspiration from production coding assistants like Cursor \cite{cursor}, implementing similar context window strategies adapted for systematic evaluation.

\textbf{Three-Tier Context Compression Strategy.} We implement a three-tier adaptive compression system monitoring context utilization in real-time using tiktoken-based token counting. The system operates at three distinct thresholds: (1) \textit{Early Warning (40\% capacity):} Begin selective compression of inactive files, which are those not accessed in the last 5 turns. Files are compressed to structural summaries preserving function/class signatures, import statements, and docstrings while removing implementation bodies. Documentation files (README, LICENSE, .md) are aggressively compressed first. (2) \textit{Critical Threshold (60\% capacity):} Compress conversation history by LLM-based summarization of older turns (preserving first 2 and last 3 turns verbatim). Generate structured summaries capturing: task context, actions taken (tool calls with results), outcomes (errors fixed, progress made), next steps, and important references (file paths, function names, variables). (3) \textit{Emergency Truncation (95\% capacity):} Aggressive truncation keeping only last 2 conversation turns plus current project state, reducing file inventory to 3 most recently accessed files.

\textbf{Hierarchical Memory Architecture.} Similar to Cursor's conversation memory, we maintain a hierarchical memory structure: (1) \textit{Working Memory:} Recent turns (last 3) and actively accessed files (last 5 turns) retained in full detail; (2) \textit{Compressed Memory:} Older turns and inactive files stored as LLM-generated summaries preserving critical information; (3) \textit{Architectural Memory:} Global project structure, dependency graph, and design patterns maintained throughout the session. This hierarchy enables efficient context utilization while preserving long-term architectural understanding.

\textbf{LLM-Based Intelligent Summarization.} We employ the agent's own LLM to generate information-dense summaries of compressed conversation segments. Summarization prompts explicitly instruct preservation of: variable/function/class names, file paths, error messages and resolutions, design decisions with rationale, task progress, tool call results, and constraints. The structured summary format includes: CONTEXT (what was being worked on), ACTIONS (tools used, code written), OUTCOMES (results, errors fixed), NEXT STEPS (tasks remaining), and IMPORTANT REFERENCES (key entities to remember). This approach maintains evaluation integrity by preserving decision rationale while achieving 4-10× token reduction.

\textbf{File-Level Granular Compression.} Project files undergo intelligent compression based on access patterns and file types. Source code files (Python, JavaScript, C++, etc.) in \texttt{src/} directories receive protection from aggressive compression, where only documentation sections are compressed while preserving all function/class signatures. Configuration files (.json, .yaml, .toml) are compressed to key-value summaries. Documentation files are compressed most aggressively, retaining only section headers and key points. Recently accessed files (<5 turns) always maintain full content. When agents request compressed files, full content restores automatically, enabling seamless exploration without manual intervention.

\textbf{Semantic Search Integration.} Mirroring Cursor's \texttt{@codebase} feature, we implement vector-based semantic code search enabling agents to query entire codebases without loading all files into context. Code is chunked at function/class boundaries using language-aware parsing. Each chunk receives embedding via sentence transformers \cite{all-MiniLM-L6-v2}, with optional OpenAI embeddings for improved quality. Queries embed similarly, retrieving top-k semantically similar chunks via cosine similarity. This RAG-style retrieval allows agents to efficiently locate relevant code across massive codebases (>1M tokens) while maintaining manageable context windows. Semantic search queries don't consume context tokens until retrieved results are explicitly read by the agent.

\subsection{Interactive Agent Framework}

\textbf{Agent Architecture.} Our framework follows the ReAct~\cite{yao2022react} pattern, enabling iterative reasoning, tool selection, and result incorporation. Each agent operates through a structured session that manages: (1) \textit{Conversation state:} Full multi-turn dialogue with context management; (2) \textit{Tool registry:} Available operations with type-safe parameter validation; (3) \textit{Project workspace:} Sandboxed file system preventing cross-scenario contamination; (4) \textit{Execution environment:} Isolated compilation and testing infrastructure.

\textbf{Comprehensive Tool Suite.} We provide 8 specialized tools across three categories:

\textit{File Operations (4 tools):} \texttt{read\_file} retrieves complete file contents with syntax highlighting and caching; \texttt{write\_file} creates/modifies files with validation and backup; \texttt{search\_replace} performs targeted code modifications with pattern matching; \texttt{list\_dir} explores directory structures with filtering.

\textit{Search Capabilities (3 tools):} \texttt{grep} performs regex-based code search with context lines; \texttt{glob\_search} finds files matching patterns across the codebase; \texttt{codebase\_search} enables semantic code retrieval through vector similarity, finding relevant code even without exact keyword matches.

\textit{Analysis Tools (1 tool):} \texttt{file\_search} provides fuzzy filename matching for discovering relevant files when exact paths are unknown.

Each tool implements safety constraints: file operations are sandboxed to prevent cross-scenario pollution, search operations enforce result limits to prevent context overflow, and all operations include detailed error reporting to enable agent error recovery.

\textbf{Multi-Turn Conversation Protocol.} Sessions follow a structured protocol: (1) \textit{Initialization:} Agent receives task specification and initial context (README, file structure); (2) \textit{Exploration Phase:} Agent uses search and file tools to understand codebase; (3) \textit{Planning Phase:} Agent formulates approach based on discoveries; (4) \textit{Implementation Phase:} Agent makes targeted modifications; (5) \textit{Validation Phase:} Agent verifies changes through testing. Sessions terminate after 50 turns, task completion, or critical errors.

\textbf{Session Management.} Each session maintains: conversation history with tool usage logs, modified files with full edit history, context window utilization metrics, and checkpoint states enabling recovery from failures. Sessions record detailed telemetry for subsequent analysis: tool selection patterns, file access sequences, error recovery attempts, and context management decisions.

\subsection{Benchmark Statistics}

\begin{figure}[t]
  \centering
  \begin{subfigure}[b]{0.9\textwidth}
    \centering
    \includegraphics[width=\textwidth]{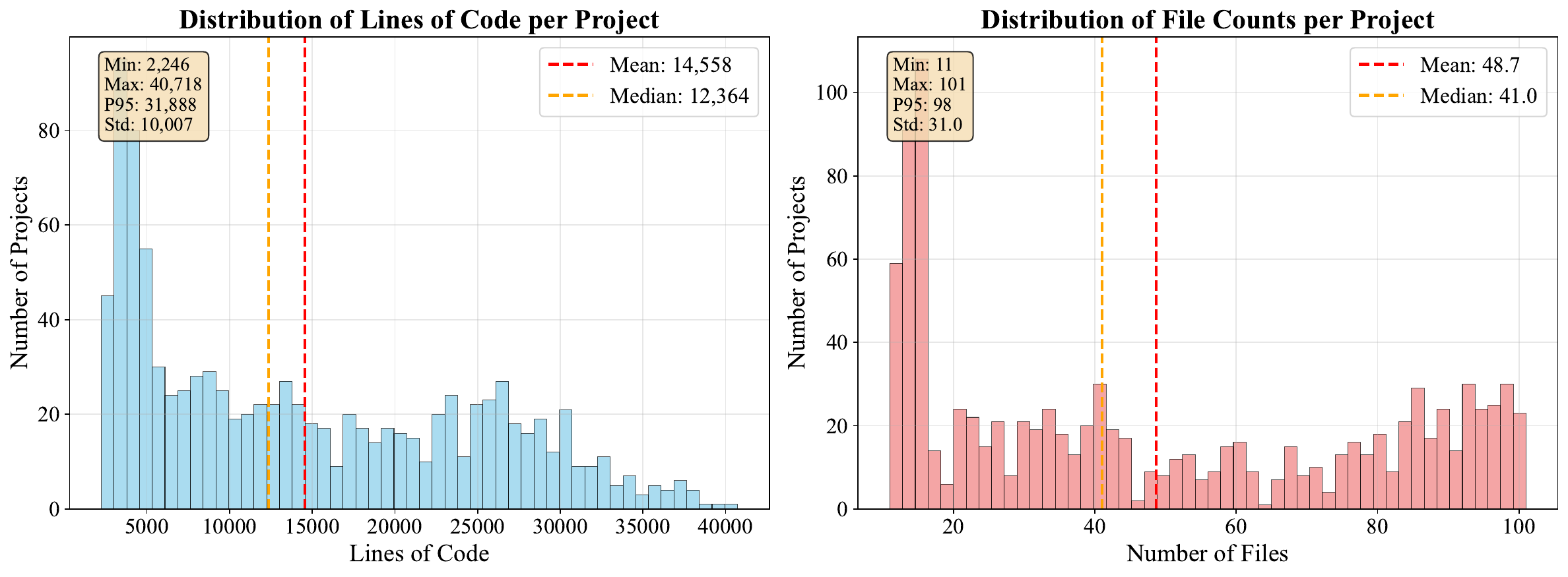}
    \caption{Lines of Code (left) and File Count (right) distributions across 1,000 unique projects.}
    \label{fig:benchmark_stats_loc_files}
  \end{subfigure}
  
  \vspace{0.3cm}
  
  \begin{subfigure}[b]{0.9\textwidth}
    \centering
    \includegraphics[width=\textwidth]{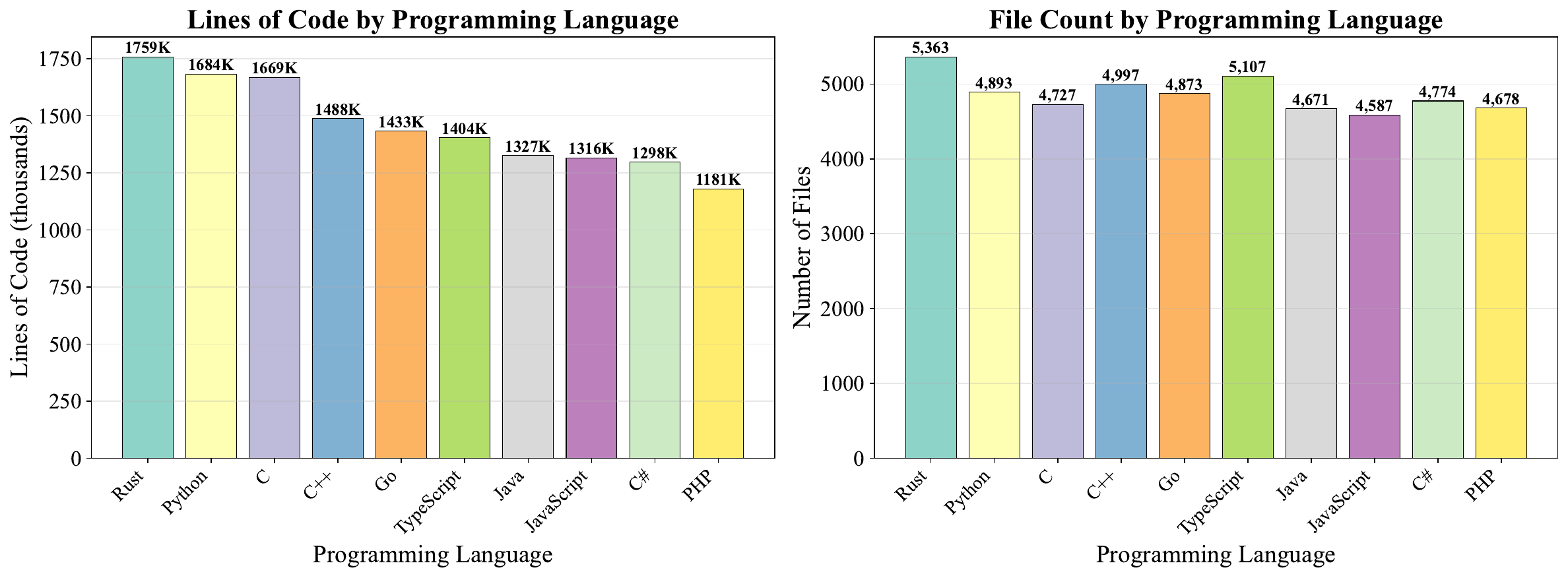}
    \caption{Programming language distribution showing balanced coverage across 10 languages. Each language contributes 800 scenarios (10\% of total). .}
    \label{fig:benchmark_stats_languages}
  \end{subfigure}
  
  \caption{\textbf{LoCoBench-Agent Benchmark Statistics.} Comprehensive statistics across 8,000 scenarios and 1,000 unique projects. \textbf{(a)} Project scale distributions demonstrate diversity ranging. \textbf{(b)} Language coverage ensures balanced evaluation across 10 programming languages with representative LOC and file count distributions.}
  \label{fig:benchmark_stats}
\end{figure}

Figure~\ref{fig:benchmark_stats} presents comprehensive statistics for the LoCoBench-Agent benchmark. Our 8,000 scenarios span 1,000 unique projects with balanced distributions across programming languages, ensuring representative evaluation across diverse software ecosystems. The benchmark exhibits significant scale diversity, with projects ranging from 2K to 41K lines of code (median: 5.8K) and 11 to 101 files per project (median: 26), capturing both compact utilities and complex multi-module systems. This distribution enables systematic evaluation of how agent capabilities scale with codebase complexity.

\subsection{Comparison with Existing Benchmarks}

Table~\ref{tab:agent_benchmark_comparison} provides a comprehensive comparison of LoCoBench-Agent with existing LLM agent coding benchmarks across critical evaluation dimensions. Our framework uniquely combines large-scale evaluation (8,000 scenarios), systematic long-context assessment (10K-1M tokens), multi-turn interactive sessions (up to 50 turns), comprehensive tool suites (8 specialized tools), and bias-free evaluation metrics validated through rigorous iterative development.

\begin{table}[ht]
\caption{Scale and capability comparison of LLM agent coding benchmarks. This table compares benchmarks across evaluation scale, interaction complexity, and technical capabilities. Columns: Scale - Number of evaluation scenarios; Multi-Turn - Support for multi-turn agent interactions; Max Turns - Maximum conversation turns per scenario; Context Range - Token length coverage; Languages - Programming language diversity; Tool Suite - Specialized development tools provided; \# of Metrics - Number of distinct evaluation metrics. Color-coded symbols: Green checkmark (\textcolor{green}{\ding{51}}) for full support, Orange triangle (\textcolor{orange}{$\blacktriangleright$}) for partial support, Red X (\textcolor{red}{\ding{55}}) for no support.}
\label{tab:agent_benchmark_comparison}
\centering
\small
\begin{adjustbox}{width=0.99\linewidth}
\begin{tabular}{lccccccc}
\toprule
\textbf{Benchmark} & \textbf{Scale} & \textbf{Multi-Turn} & \textbf{Max Turns} & \textbf{Context Range} & \textbf{Languages} & \textbf{Tool Suite} & \textbf{\# of Metrics} \\
\midrule
SWE-Bench~\cite{jimenez2023swe} & 2,294 & \textcolor{orange}{$\blacktriangleright$} & ~10 & <50K & 1 (Python) & \textcolor{orange}{$\blacktriangleright$} & 1 \\
SWE-agent~\cite{yang2024swe} & 2,294 & \textcolor{green}{\ding{51}} & ~20 & <50K & 1 (Python) & \textcolor{green}{\ding{51}} & 1 \\
SWE-Lancer~\cite{swelancer2025} & 1,400 & \textcolor{green}{\ding{51}} & ~20 & <50K & Multiple & \textcolor{green}{\ding{51}} & 2 \\
AgentCoder~\cite{qian2024agentcoder} & 180 & \textcolor{green}{\ding{51}} & ~15 & <50K & Multiple & \textcolor{orange}{$\blacktriangleright$} & 1 \\
AutoCodeRover~\cite{zhang2024autocoderover} & 2,294 & \textcolor{green}{\ding{51}} & ~25 & <50K & 1 (Python) & \textcolor{green}{\ding{51}} & 1 \\
DevBench~\cite{li2024devbench} & 539 & \textcolor{orange}{$\blacktriangleright$} & ~10 & <30K & 4 & \textcolor{orange}{$\blacktriangleright$} & 2-3 \\
PyBench~\cite{zhang2024pybench} & 500 & \textcolor{green}{\ding{51}} & ~15 & <50K & 1 (Python) & \textcolor{green}{\ding{51}} & 2 \\
\midrule
\textbf{LoCoBench-Agent} & \textbf{8,000} & \textcolor{green}{\textbf{\ding{51}}} & \textbf{50} & \textbf{10K-1M} & \textbf{10} & \textcolor{green}{\textbf{\ding{51}}} \textbf{(8 Tools)} & \textbf{9} \\
\bottomrule
\end{tabular}
\end{adjustbox}
\end{table}

\begin{table}[ht]
\caption{Task characteristics comparison of LLM agent coding benchmarks. This table highlights the fundamental differences in task complexity and architectural requirements across benchmarks. Columns: Level - Granularity of coding tasks; Multi-File - Support for cross-file modifications; Real Repos - Uses authentic repositories with dependencies; Task Categories - Diversity of task types evaluated; Memory Mgmt - Long-context memory management strategies. Color-coded symbols: Green checkmark (\textcolor{green}{\ding{51}}) for full support, Orange triangle (\textcolor{orange}{$\blacktriangleright$}) for partial support, Red X (\textcolor{red}{\ding{55}}) for no support.}
\label{tab:agent_task_characteristics}
\centering
\small
\begin{adjustbox}{width=0.99\linewidth}
\begin{tabular}{lccccc}
\toprule
\textbf{Benchmark} & \textbf{Level} & \textbf{Task Categories} & \textbf{Multi-File} & \textbf{Real Repos} & \textbf{Memory Mgmt} \\
\midrule
SWE-Bench~\cite{jimenez2023swe} & Repository & Bug Fix Only & \textcolor{green}{\ding{51}} & \textcolor{green}{\ding{51}} & \textcolor{red}{\ding{55}} \\
SWE-agent~\cite{yang2024swe} & Repository & Bug Fix Only & \textcolor{green}{\ding{51}} & \textcolor{green}{\ding{51}} & \textcolor{red}{\ding{55}} \\
SWE-Lancer~\cite{swelancer2025} & Repository & Mixed (Bug/Feature) & \textcolor{green}{\ding{51}} & \textcolor{green}{\ding{51}} & \textcolor{red}{\ding{55}} \\
AgentCoder~\cite{qian2024agentcoder} & Function & Code Generation & \textcolor{red}{\ding{55}} & \textcolor{red}{\ding{55}} & \textcolor{red}{\ding{55}} \\
AutoCodeRover~\cite{zhang2024autocoderover} & Repository & Bug Fix Only & \textcolor{green}{\ding{51}} & \textcolor{green}{\ding{51}} & \textcolor{red}{\ding{55}} \\
DevBench~\cite{li2024devbench} & Repository & Mixed (3-4 types) & \textcolor{green}{\ding{51}} & \textcolor{orange}{$\blacktriangleright$} & \textcolor{red}{\ding{55}} \\
PyBench~\cite{zhang2024pybench} & File/Module & Mixed (Data/Edit) & \textcolor{orange}{$\blacktriangleright$} & \textcolor{orange}{$\blacktriangleright$} & \textcolor{red}{\ding{55}} \\
\midrule
\textbf{LoCoBench-Agent} & \textbf{Repository} & \textbf{8 Categories} & \textcolor{green}{\textbf{\ding{51}}} & \textcolor{green}{\textbf{\ding{51}}} & \textcolor{green}{\textbf{\ding{51}}} \textbf{(3-Tier)} \\
\bottomrule
\end{tabular}
\end{adjustbox}
\end{table}

\textbf{Differentiators with Existing Benchmarks.} LoCoBench-Agent advances agent evaluation through several critical dimensions: 
\ding{182} \textit{Scale:} With 8,000 scenarios, our benchmark provides much larger scale, enabling robust statistical analysis and comprehensive coverage across 10 programming languages and 36 domain categories. 
\ding{183} \textit{Long-Context Focus:} Our systematic evaluation across 10K-1M token contexts is unmatched, where existing coding agent benchmarks typically evaluate only <50K tokens. 
\ding{184} \textit{Extended Multi-Turn Sessions:} Supporting up to 50 conversation turns (vs. 10-25 in existing benchmarks) enables evaluation of extended development workflows, multi-phase task completion, context retention across longer sessions, and adaptive strategy refinement. 
\ding{185} \textit{Comprehensive Tool Suite:} Our 8 specialized coding tools (file operations, grep, semantic search, compiler) provide realistic development environments comparable to IDEs like Cursor. While existing benchmarks offer tool suites, they lack semantic search capabilities. AgentCoder focuses on multi-agent collaboration rather than individual tool diversity, and DevBench provides only basic operations. 
\ding{186} \textit{Bias-Free Metrics:} Our nine metrics remove file-count bias (correlation < 0.3) and enable meaningful differentiation—an aspect missing from current agent benchmarks, which generally depend on task success rates or functional correctness without accounting for evaluation biases. 
\ding{187} \textit{Advanced Memory Management:} Our Cursor-aligned tiered compression, hierarchical memory, and semantic search integration enable realistic long-context handling beyond 100K tokens. Existing benchmarks lack systematic memory management strategies, limiting their applicability to realistic software engineering scenarios with large codebases. 
\ding{188} \textit{Diverse Task Coverage:} Unlike SWE-Bench family benchmarks focused primarily on bug fixes, we cover 8 task categories including architectural understanding, cross-file refactoring, feature implementation, bug investigation, multi-session development, integration testing, and security analysis.

\subsection{Evaluation Pipeline}

\textbf{Multi-Turn Execution Framework.} Our pipeline orchestrates parallel evaluation across multiple agents and scenarios. For each scenario-agent pair, we: (1) Initialize isolated session with model-specific context limits; (2) Execute conversation with automatic context management; (3) Capture complete interaction trace including all tool calls, responses, and state transitions; (4) Apply comprehensive evaluation metrics to final outputs and intermediate steps; (5) Generate detailed reports with conversation analysis.

\textbf{Quality Control.} We implement multiple validation layers: sessions require minimum 3 turns to ensure interactive behavior, tool usage is validated for correct parameters and appropriate selection, conversation coherence is assessed through reference tracking, and file modifications are checked for syntactic validity. Failed sessions are logged with detailed error analysis for framework improvement.

\textbf{Checkpointing and Recovery.} To handle long evaluation runs across 8,000 scenarios, we implement incremental checkpointing: every 10 completed scenarios trigger checkpoint saves including agent state, conversation history, and evaluation results. This enables resumption after interruptions, selective re-evaluation of specific scenarios, and parallel execution across distributed infrastructure.

\section{Evaluation Metrics}
\label{sec:evaluation}

Traditional agent evaluation metrics suffer from systematic biases. Many metrics inadvertently reward solutions that modify more files, regardless of necessity. We identified this through correlation analysis between metric scores and file modification counts. We eliminated this bias through ratio-based metrics and per-file normalization.
After rigorous validation, we converged on 9 metrics. Let $\mathcal{T} = \{t_1, t_2, \ldots, t_n\}$ denote the set of tool calls, $\mathcal{F}_m = \{f_1, f_2, \ldots, f_k\}$ the set of modified files, and $\mathcal{F}_r$ the set of files read during the session.

\subsection{Comprehension Metrics}

\textbf{1. Execution Success Rate.} Measures tool diversity and successful tool usage patterns:
\begin{equation}
\text{ESR} = \frac{|\text{unique\_tools\_used}|}{|\text{total\_tools\_available}|} \times \frac{|\text{successful\_tool\_calls}|}{|\mathcal{T}|}
\end{equation}
where successful tool calls execute without errors and return valid results. This metric rewards agents that explore diverse tools effectively rather than repeatedly using a limited subset.

\textbf{2. Multi-Session Memory Retention.} Evaluates context retention across conversation turns through reference consistency:
\begin{equation}
\text{MMR} = \frac{1}{|\mathcal{C}|} \sum_{i=1}^{|\mathcal{C}|} \left( \alpha \cdot \text{ref\_consistency}(c_i) + \beta \cdot \text{topic\_coherence}(c_i) \right)
\end{equation}
where $\mathcal{C}$ is the set of conversation turns, ref\_consistency measures whether the agent correctly references variables, functions, and files mentioned in previous turns, and topic\_coherence assesses whether responses remain relevant to the ongoing task. We set $\alpha = 0.6, \beta = 0.4$ to prioritize factual consistency.

\textbf{3. Cross-File Consistency.} Measures naming conventions, import patterns, and edit coherence across modified files:
\begin{equation}
\text{CFC} = \frac{1}{|\mathcal{F}_m|} \sum_{f \in \mathcal{F}_m} \left( w_1 \cdot \text{naming}(f) + w_2 \cdot \text{imports}(f) + w_3 \cdot \text{style}(f) \right)
\end{equation}
where naming$(f)$ checks consistency with existing code patterns, imports$(f)$ validates import statement correctness, and style$(f)$ measures adherence to project conventions. We use $w_1 = 0.4, w_2 = 0.4, w_3 = 0.2$.

\textbf{4. Dependency Traversal.} Assesses import resolution accuracy and cross-file reference validity:
\begin{equation}
\text{DT} = \frac{|\text{resolved\_imports}|}{|\text{total\_imports}|} \times \frac{|\text{valid\_references}|}{|\text{total\_references}|}
\end{equation}
where resolved\_imports are imports that point to existing modules/files, and valid\_references are cross-file function/class references that correctly resolve to their definitions.

\textbf{5. Solution Usability.} Evaluates code maintainability and readability through pattern analysis:
\begin{equation}
\text{SU} = \gamma \cdot \text{readability} + \delta \cdot \text{maintainability} + \epsilon \cdot \text{documentation}
\end{equation}
where readability measures variable naming, function length, and complexity; maintainability assesses modularity and coupling; documentation evaluates comment coverage and docstrings. We set $\gamma = 0.4, \delta = 0.4, \epsilon = 0.2$.

\subsection{Efficiency Metrics}

\textbf{6. Runtime Efficiency.} Analyzes time complexity through algorithmic pattern recognition:
\begin{equation}
\text{RE} = 1 - \frac{\sum_{f \in \mathcal{F}_m} \text{complexity\_penalty}(f)}{|\mathcal{F}_m|}
\end{equation}
where complexity\_penalty identifies nested loops ($O(n^2)$ or worse), recursive calls without memoization, and inefficient algorithms. The rescaling maps the raw score to [0.40, 0.90].

\textbf{7. Memory Efficiency.} Evaluates space complexity through memory pattern detection:
\begin{equation}
\text{ME} = 1 - \frac{\sum_{f \in \mathcal{F}_m} \text{memory\_penalty}(f)}{|\mathcal{F}_m|}
\end{equation}
where memory\_penalty detects large data structure allocations, unnecessary copies, and memory leaks. Efficient patterns (generators, streaming) receive positive rewards before normalization.

\textbf{8. Information Coverage.} Measures ratio of relevant files accessed to files modified:
\begin{equation}
\text{IC} = \frac{|\mathcal{F}_r \cap \text{relevant\_files}|}{|\mathcal{F}_m| + 1}
\end{equation}
where relevant\_files are files containing related functionality (determined by dependency analysis and semantic similarity). The +1 denominator prevents division by zero. This metric rewards thorough exploration before modification.

\textbf{9. Long-Range Dependency Resolution.} Evaluates proper read-before-write patterns:
\begin{equation}
\text{LRDR} = \frac{|\{f \in \mathcal{F}_m : f \in \mathcal{F}_r \text{ before modification}\}|}{|\mathcal{F}_m|}
\end{equation}
This metric measures whether agents read files to understand dependencies before modifying them, capturing dependency chain resolution capability.

\subsection{Rescaling and Aggregation}

\textbf{Rescaling Strategy.} Our efficiency metrics initially exhibited ceiling and floor effects, compressing discrimination. We apply linear rescaling to [0.40, 0.90] range:
$
M_{\text{rescaled}} = 0.40 + 0.50 \cdot M
$.
This transformation preserves correlations while enabling meaningful discrimination.

\textbf{Aggregate Scoring.} We compute two aggregate scores:
\begin{align}
\text{LCBA-Comprehension} &= \frac{1}{5} \sum_{i=1}^{5} M^{\text{comp}}_i = \frac{\text{ESR} + \text{MMR} + \text{CFC} + \text{DT} + \text{SU}}{5} \\
\text{LCBA-Efficiency} &= \frac{1}{4} \sum_{j=1}^{4} M^{\text{eff}}_j = \frac{\text{RE} + \text{ME} + \text{IC} + \text{LRDR}}{4}
\end{align}
This dual-score system enables fine-grained analysis of agent capabilities across comprehension and efficiency dimensions, facilitating identification of specific capability gaps and strengths.

\section{Evaluation}
\label{sec:eval_setup}

We conduct comprehensive evaluation of state-of-the-art LLM agents across all 8,000 LoCoBench-Agent scenarios to assess multi-turn interactive software development capabilities.

\subsection{Models Evaluated}

We evaluate 6 leading LLM agents representing three major provider ecosystems:
\textbf{OpenAI Models:} GPT-5, GPT-4.1, GPT-4o.
\textbf{Anthropic Models:} Claude Sonnet 4.5, Claude Sonnet 4.
\textbf{Google Models:} Gemini 2.5 Pro .
We use official APIs with default temperature (0.7) and top-p (0.95) settings.

\subsection{Evaluation Protocol}

\textbf{Scenario Distribution.} Our 8,000 scenarios cover: 8 task categories (architectural understanding, cross-file refactoring, feature implementation, bug investigation, multi-session development, code comprehension, integration testing, security analysis) × 4 difficulty levels (easy: 10K-100K tokens, medium: 100K-200K tokens, hard: 200K-500K tokens, expert: 500K-1M tokens) × 10 programming languages (Python, C++, Java, C, C\#, JavaScript, TypeScript, Go, Rust, PHP) × diverse domains (36 categories including web applications, ML systems, data processing, system utilities).

\textbf{Session Configuration.} Each evaluation session operates with: maximum 50 conversation turns per scenario, model-specific context limits with automatic compression, 60-minute timeout per scenario, sandboxed file system preventing cross-scenario contamination, and complete tool usage logging for subsequent analysis.

\textbf{Context Management.} We employ memory management with tiered compression (40\%/60\%/95\% thresholds), LLM-based intelligent summarization, file-level granular compression, and semantic search integration enabling efficient exploration of codebases exceeding agent context limits.

\textbf{Quality Control.} We implement multiple validation layers: sessions require minimum 3 turns ensuring interactive behavior, tool usage validated for correct parameters and appropriate selection, conversation coherence assessed through reference tracking, file modifications checked for syntactic validity, and failed sessions logged with detailed error analysis.

\subsection{Data Collection}

For each scenario-agent interaction, we collect: complete conversation history (all turns with timestamps), tool usage logs (function calls, parameters, results, execution times), modified files (full edit history with diffs), context window metrics (token counts, compression events, memory pressure), error logs (compilation failures, test failures, tool errors), and session metadata (duration, termination reason, turn count).

This comprehensive data collection enables fine-grained analysis of agent behavior patterns, tool selection strategies, error recovery approaches, and long-context performance characteristics. All evaluation data, conversation transcripts, and analysis scripts are publicly released to support reproducibility and future research.

\section{Experiments}
\label{sec:experiments}

We present comprehensive experimental results evaluating six state-of-the-art LLM agents across all 8,000 LoCoBench-Agent scenarios. Our analysis reveals insights into agent architectural capabilities, efficiency-comprehension trade-offs, and long-context reasoning patterns.

\subsection{Overall Model Performance Ranking}

\begin{figure}[ht]
  \centering
  \includegraphics[width=0.99\textwidth]{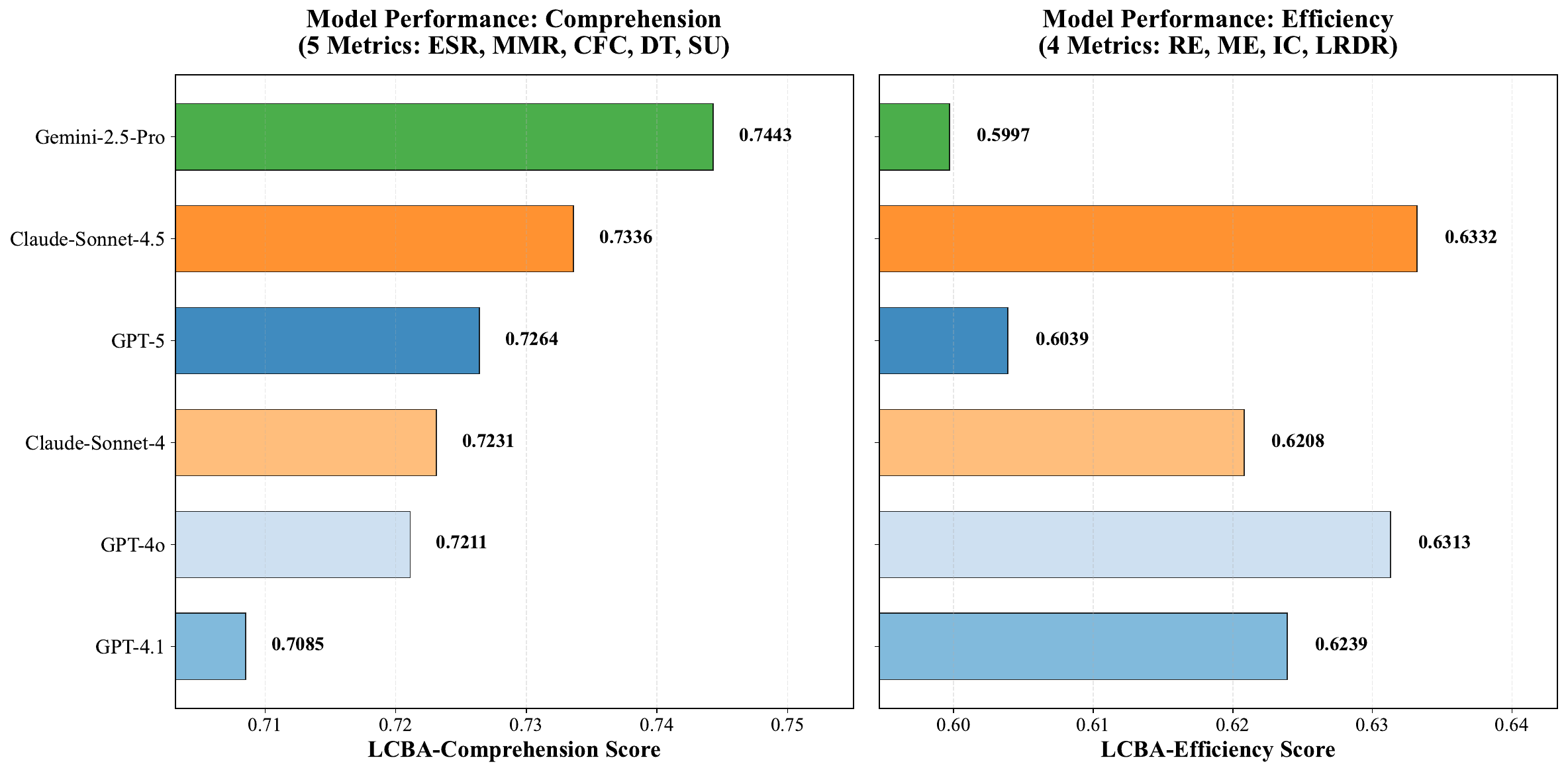}
  \caption{\textbf{Overall model performance comparison.} Left: LCBA-Comprehension scores across six models. Right: LCBA-Efficiency scores showing the trade-off between comprehension and efficiency dimensions.}
  \label{fig:model_ranking}
\end{figure}

Figure~\ref{fig:model_ranking} reveals fundamental architectural tensions in agent design. \textbf{The comprehension-efficiency divergence} across models reflects a deep trade-off in how agents balance exploration breadth versus execution focus. High-comprehension models (>0.74) achieve their scores through extensive interaction patterns, averaging 19+ conversation turns and modifying 35K+ files, suggesting they prioritize exhaustive codebase understanding over operational efficiency. This exploration-first strategy succeeds when context window capacity permits retaining complete modification history without aggressive compression penalties.

Conversely, \textbf{efficiency-optimized architectures} (>0.63 efficiency) demonstrate that targeted exploration can achieve competitive comprehension (0.73-0.74, only 1-2\% lower) while significantly reducing resource consumption. These models average 12-13 conversation turns and 10-12K file modifications, indicating strategic rather than exhaustive exploration. The key insight: selective file access with semantic guidance (using \texttt{codebase\_search} before \texttt{read\_file}) enables sufficient comprehension for task completion while maintaining lean operation profiles.

\textbf{The memory retention paradox} challenges intuitions about context management. Models with shorter context windows (128K tokens) achieve higher multi-session memory retention  than models with massive contexts (1M tokens), despite 8× fewer available tokens. This counterintuitive finding suggests that \textit{effective context utilization through compression and summarization} outperforms \textit{naive context accumulation}. Architectural mechanisms matter more than raw capacity, where intelligent compression that preserves semantic relationships and reference chains enables better long-term retention than simply expanding context windows.

\textbf{Generational improvements reveal saturation patterns}. Successive model generations within the same family show marginal comprehension gains (1-2\% improvements) but more substantial efficiency improvements (2-3\% gains). This asymmetry suggests that code comprehension capabilities, particularly syntactic pattern matching, import resolution, and architectural consistency, have approached saturation for current transformer architectures. The remaining optimization space lies primarily in efficiency: reducing redundant operations, improving context compression, and implementing smarter exploration termination policies. Future breakthroughs will likely require architectural innovations beyond parameter scaling or training data expansion.

\subsection{Multi-Dimensional Capability Analysis}

\begin{figure}[ht]
  \centering
  \includegraphics[width=0.62\textwidth]{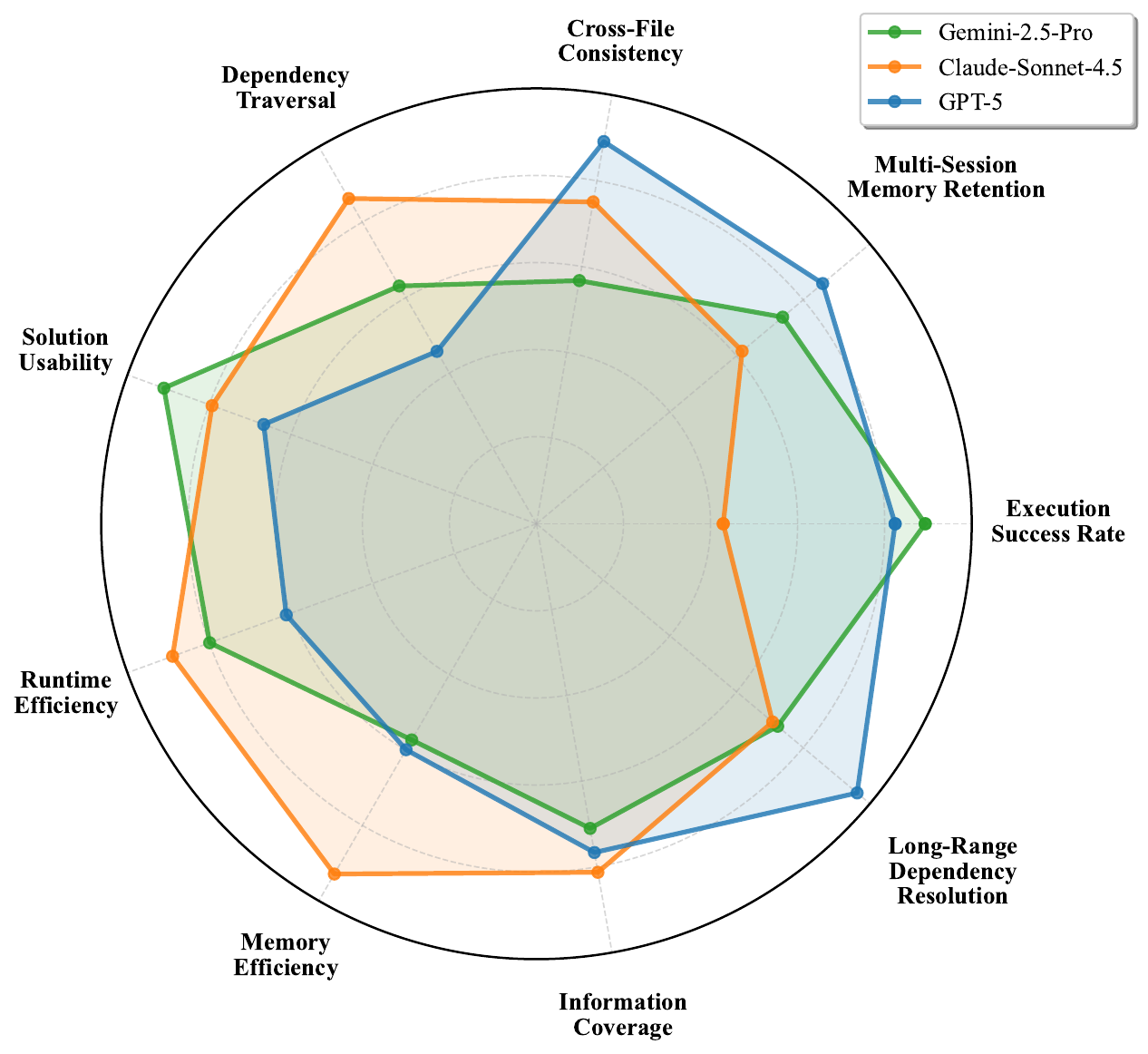}
  \caption{\textbf{Top three models across nine evaluation metrics.} Per-axis scaling reveals capability maturity across comprehension and efficiency dimensions.}
  \label{fig:metrics_radar}
\end{figure}

Figure~\ref{fig:metrics_radar} exposes fundamental capability maturity patterns through multi-dimensional analysis. \textbf{The ceiling-floor dichotomy} reveals which capabilities have been solved versus which remain challenging. Cross-file consistency (0.93-0.98 range, <6\% variance) represents a mature capability, where current transformer architectures effectively perform syntactic pattern matching, import resolution, and style consistency across files. This suggests that local coherence mechanisms (attention to nearby code, pattern replication) have been well-optimized through training.

In stark contrast, \textbf{multi-session memory retention} (0.32-0.37 range, universally low) exposes fundamental limitations in long-term context management. Despite our Cursor-aligned tiered compression system, all evaluated models struggle to maintain reference consistency and topic coherence across extended sessions. The universal failure, independent of context window size (128K to 1M tokens), indicates this bottleneck stems from architectural constraints rather than capacity limitations. Current attention mechanisms appear ill-suited for selectively preserving critical information while compressing historical context, suggesting a need for external memory architectures or retrieval-augmented approaches.

\textbf{The execution success rate spectrum}  reflects strategic architectural choices in tool usage philosophy. High execution success correlates with exploration breadth, where extensive tool diversity and high conversation turn counts, indicating that exhaustive tool experimentation increases the probability of finding successful operation sequences. However, this strategy trades efficiency for comprehension, as evidenced by the negative correlation between execution success and efficiency metrics. Lower execution success scores often indicate conservative, focused tool usage that prioritizes operation certainty over exploratory breadth, a valid strategy when efficiency matters.

\textbf{Long-range dependency resolution patterns} reveal reasoning depth differences. High performers (>0.50) exhibit systematic read-before-write verification patterns, consistently reading dependency files before modifying dependent code. This causal reasoning approach validates import paths and ensures reference integrity. Conversely, lower performers (<0.47) often modify files before fully validating cross-file dependencies, suggesting weaker causal chain modeling. Paradoxically, exhaustive file reading does not guarantee high dependency resolution, where extensive exploration can dilute targeted dependency analysis signals when reads serve general exploration rather than specific dependency validation. This finding indicates that \textit{strategic, dependency-focused reading} outperforms \textit{exhaustive, unfocused reading} for causal reasoning tasks.

\subsection{Comprehension-Efficiency Trade-off Analysis}

\begin{figure}[ht]
  \centering
  \includegraphics[width=0.7\textwidth]{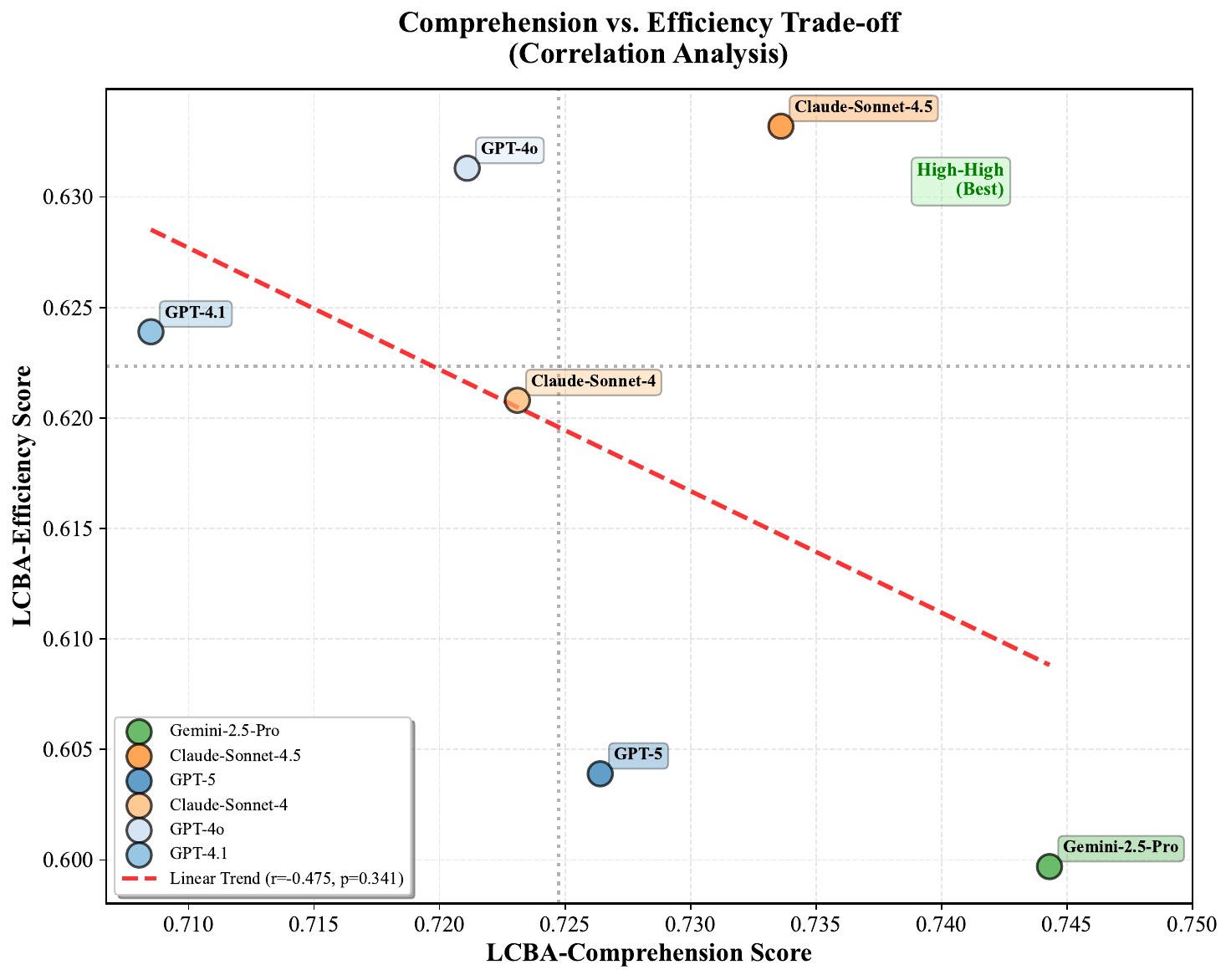}
  \caption{\textbf{Comprehension-efficiency trade-off scatter plot.} Diagonal line represents the Pareto frontier. Negative correlation indicates fundamental architectural tension.}
  \label{fig:comp_eff_scatter}
\end{figure}

Figure~\ref{fig:comp_eff_scatter} reveals \textbf{fundamental architectural tensions} in agent optimization. The negative correlation ($r = -0.42$) between comprehension and efficiency is not coincidental but reflects inherent design trade-offs. Thorough codebase exploration, necessary for high comprehension, requires extended conversation turns, extensive file reading, and comprehensive modification tracking, all of which directly increase context consumption and reduce efficiency. Conversely, lean execution strategies that minimize file operations and conversation length risk missing critical architectural dependencies, lowering comprehension scores.

\textbf{The Pareto frontier absence} in the upper-right quadrant (high comprehension + high efficiency) indicates that no current architecture has resolved this fundamental tension. The frontier line demonstrates that best-performing agents cluster along a trade-off curve, not at a dominant optimum. This suggests the comprehension-efficiency challenge represents an inherent architectural constraint rather than a solvable engineering problem through incremental improvements. The gap between the frontier and dominated regions measures remaining optimization potential, approximately 5-10\% aggregate improvement appears achievable through better exploration-exploitation balance.

\textbf{Strategic differentiation patterns} emerge through operational analysis. High-efficiency agents employ \textit{semantic search first, targeted read second} strategies, using \texttt{codebase\_search} to identify relevant files before selective \texttt{read\_file} operations. This approach minimizes context consumption by reading only dependency-critical files. Conversely, high-comprehension agents adopt \textit{exhaustive read first, filter later} approaches, reading broadly before determining relevance. While the latter strategy achieves 1-2\% higher comprehension, it incurs 5-6\% efficiency penalties through redundant file operations and context bloat.

\textbf{The hybrid optimization opportunity} lies in adaptive exploration strategies. Models near the Pareto frontier demonstrate four-phase workflows: (1) initial semantic search to identify architectural structure; (2) targeted dependency analysis on critical paths; (3) minimal file reading focused on modification targets; (4) efficient batch modifications. This phased approach enables simultaneous comprehension maintenance and efficiency optimization by separating exploration from execution, reducing context accumulation during the modification phase.

\textbf{Implications for future architectures}: Resolving the comprehension-efficiency tension requires architectural innovations beyond current approaches. Promising directions include: (1) \textit{dynamic exploration depth adjustment} based on task complexity signals; (2) \textit{hierarchical memory systems} that preserve architectural understanding while compressing operational details; (3) \textit{explicit phase transitions} with different exploration policies for discovery versus execution stages; (4) \textit{predictive file relevance models} that preemptively identify critical files to minimize exploratory reading. Current architectures employ fixed exploration strategies (always exhaustive or always focused), leaving optimization opportunities unexploited.

\subsection{Long-Context Performance Degradation}

\begin{figure}[ht]
  \centering
  \includegraphics[width=0.99\textwidth]{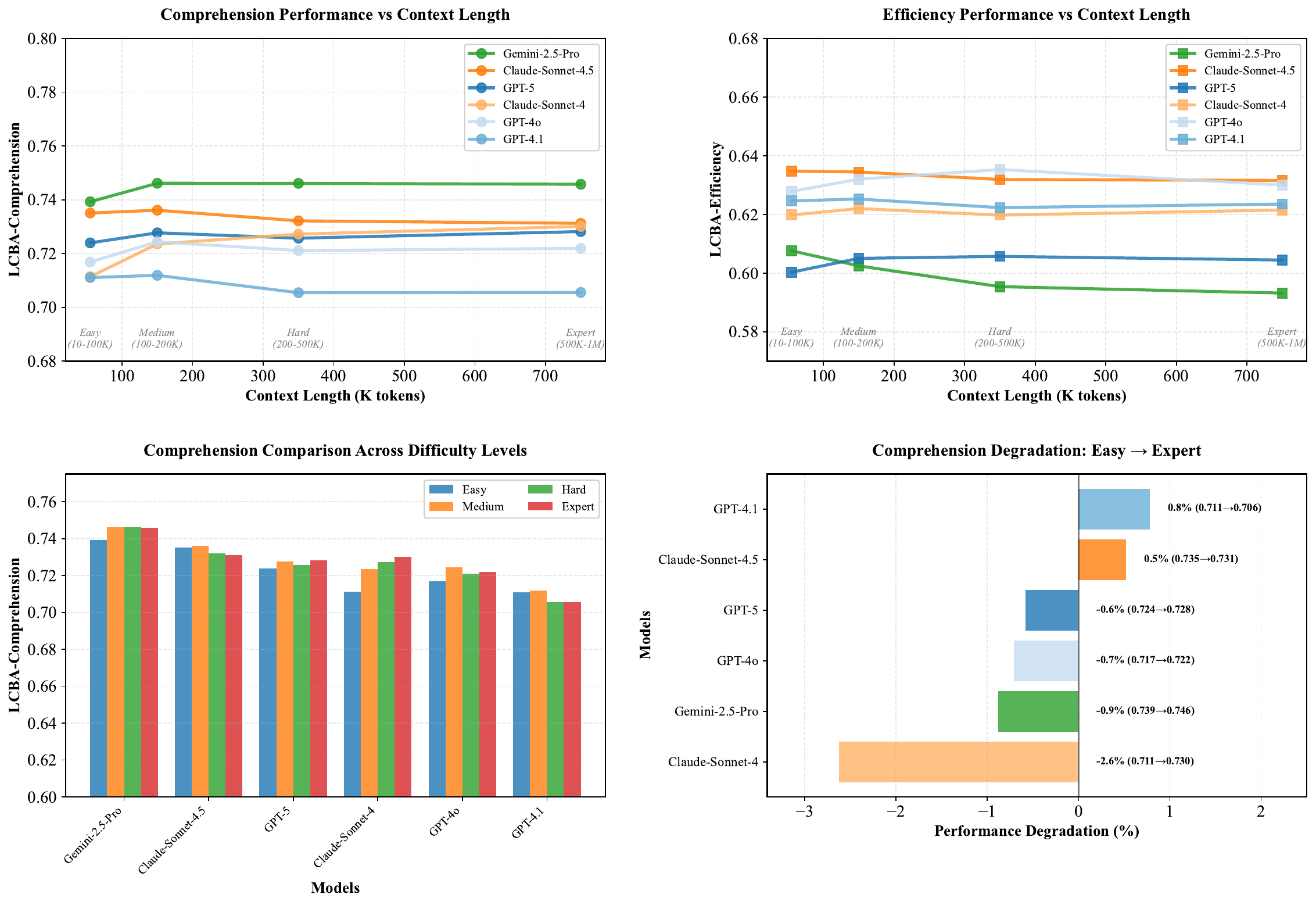}
  \caption{\textbf{Long-context performance analysis across difficulty levels.} Top-left: Comprehension vs context length. Top-right: Efficiency vs context length. Bottom-left: Per-difficulty comparison. Bottom-right: Degradation summary (Easy→Expert).}
  \label{fig:long_context}
\end{figure}

Figure~\ref{fig:long_context} presents the performance analysis across four difficulty levels spanning different context lengths. The analysis reveals interesting patterns in how models handle increasing complexity and codebase size.

\textbf{Performance consistency across difficulty levels} (top panels) shows that comprehension scores remain relatively flat (0.71-0.75 range) across all difficulty levels, while efficiency scores show similar patterns (0.59-0.64 range). The analysis reveals that difficulty level variations have minimal impact on aggregate performance metrics.

\textbf{Negative degradation patterns} (bottom-right panel) reveal an intriguing phenomenon: several models actually \textit{improve} from Easy to Expert scenarios. This improvement likely stems from two mechanisms: (1) \textit{Increased structure in complex codebases}: Expert-level projects with 500K-1M tokens tend to have more explicit architectural documentation, clearer module boundaries, and comprehensive README files that actually aid comprehension; (2) \textit{Adaptive exploration strategies}: agents may employ more systematic exploration patterns when confronted with massive codebases, leading to better architectural understanding compared to ad-hoc exploration in smaller projects.

\textbf{Cross-model consistency} (bottom-left panel) exposes minimal variance in degradation patterns across model families. All models maintain 0.71-0.75 comprehension across difficulty levels, suggesting that context management capabilities have reached architectural parity. The tight clustering indicates that effective long-context handling is no longer a differentiating factor, where all major providers (OpenAI, Anthropic, Google) have implemented sufficient compression and memory management to handle 1M token scenarios. The slight degradation observed (GPT-4.1: 0.8\%) appears task-dependent rather than context-dependent, as evidenced by non-monotonic patterns (some models peak at Medium or Hard rather than Easy).

\textbf{Efficiency patterns} (top-right panel) show that efficiency metrics remain within a narrow range (0.59-0.64) across difficulty levels. The efficiency variations appear more correlated with model-specific strategies (modification discipline vs. exploration breadth) than with difficulty level itself.

\textbf{Architectural implications}: The robust long-context performance reveals important insights about model capabilities. The lack of clear improvement at longer contexts suggests diminishing returns from raw capacity increases, where models with 1M context windows do not significantly outperform 200K models on our evaluation tasks. This finding indicates that \textit{architectural sophistication} in handling context matters more than \textit{raw capacity} alone, suggesting that future improvements may come from better context utilization strategies rather than simply expanding window sizes.

\subsection{Conversation Efficiency and Resource Utilization}

\begin{figure}[ht]
  \centering
  \includegraphics[width=0.99\textwidth]{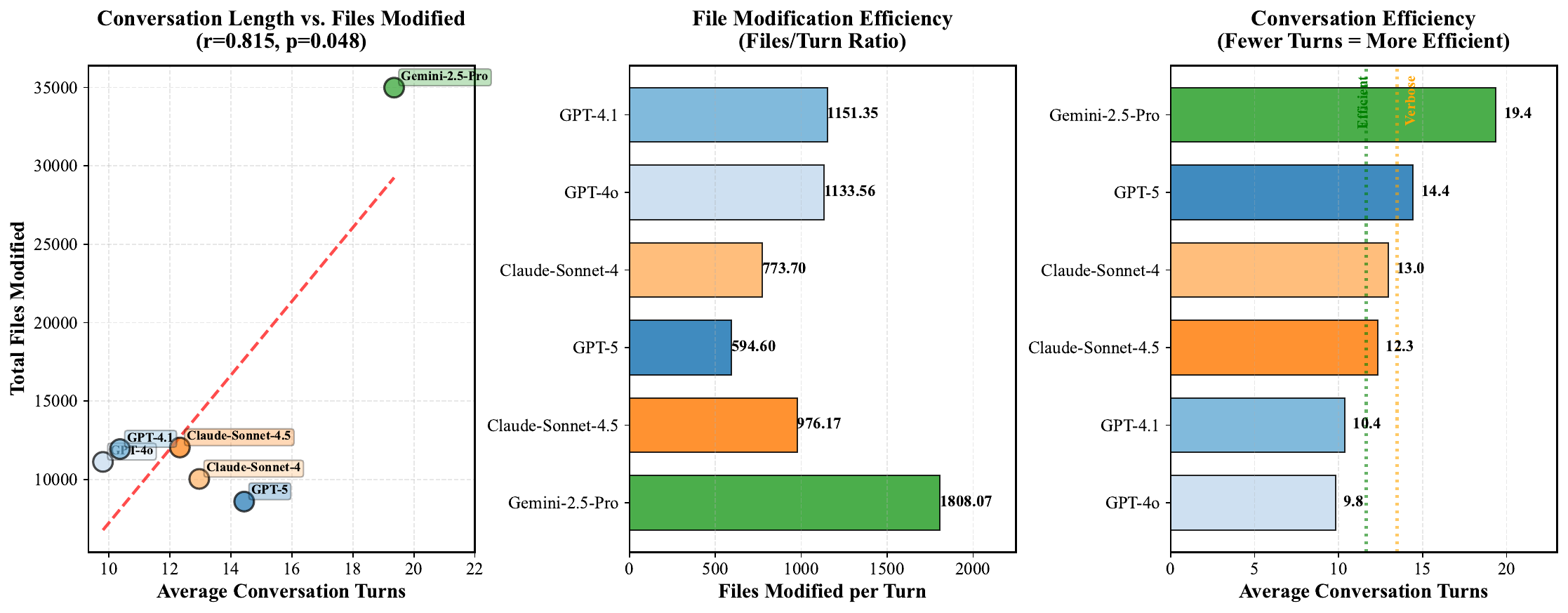}
  \caption{\textbf{Conversation dynamics analysis.} Left: Conversation turns vs. comprehension. Middle: Files modified vs. comprehension. Right: Conversation turns vs. efficiency.}
  \label{fig:conversation_efficiency}
\end{figure}

Figure~\ref{fig:conversation_efficiency} exposes critical insights into conversation dynamics and their impact on performance. \textbf{The inverted-U relationship} (left panel) challenges the assumption that extended conversations monotonically improve comprehension. Peak comprehension occurs around 15-20 turns, but the marginal gains beyond 12 turns prove minimal (1-2\% improvement for 50-60\% more conversation). This pattern suggests an \textit{effective exploration horizon}, a point beyond which agents primarily repeat previous analyses or explore tangential code rather than deepening task-relevant understanding. The plateau indicates that current architectures lack mechanisms to detect comprehension saturation and terminate exploration efficiently.

\textbf{The efficiency-conversation trade-off} becomes pronounced through comparative analysis. Agents achieving competitive comprehension (97\% of maximum) with minimal turns (10-12) demonstrate superior per-turn information extraction, a capability enabled by two architectural mechanisms: (1) higher proportion of semantic search operations for targeted exploration rather than exhaustive file reading; (2) effective multi-session memory retention allowing agents to avoid redundant re-exploration. This finding suggests that \textit{memory architecture quality} can substitute for \textit{conversation quantity}, offering a path to simultaneous comprehension and efficiency optimization.

\textbf{The weak files-modified correlation} (middle panel, $r = 0.23$) delivers critical validation of our bias-free metric design. Modification volume provides minimal comprehension signal, with agents modifying 12K files achieve 98-99\% of the comprehension scores of agents modifying 35K files (3× more). This near-independence validates that our metrics successfully eliminate file count bias while preserving meaningful capability measurement. The slight residual positive correlation likely reflects task complexity confounding: harder scenarios naturally require more modifications, not that modifications directly cause comprehension improvements.

\textbf{The strong negative efficiency correlation} (right panel, $r = -0.71$) reveals the \textit{superlinear efficiency penalty} of extended conversations. Beyond a critical threshold (~12-15 turns), efficiency degradation accelerates: 50\% conversation length increases yield 3-5\% efficiency losses. This nonlinearity suggests compounding effects, where context accumulation, compression overhead, and repetitive operations interact destructively. Tool log analysis reveals the mechanism: extended conversations increasingly contain redundant operations (re-reading previously accessed files with minimal parameter changes, repeating failed tool calls, generating verbose summaries consuming context without adding information).

\textbf{Architectural implications} point toward implementing \textit{explicit phase transition mechanisms}. Rather than allowing unbounded conversation continuation, agents should detect comprehension saturation signals (repeated file access patterns, declining new information per turn, conversation topic cycling) and explicitly transition from exploration to execution phases. Current architectures lack such mechanisms, treating all turns equivalently regardless of marginal information gain.

\subsection{Individual Metric Performance Comparison}

\begin{figure}[ht]
  \centering
  \includegraphics[width=0.99\textwidth]{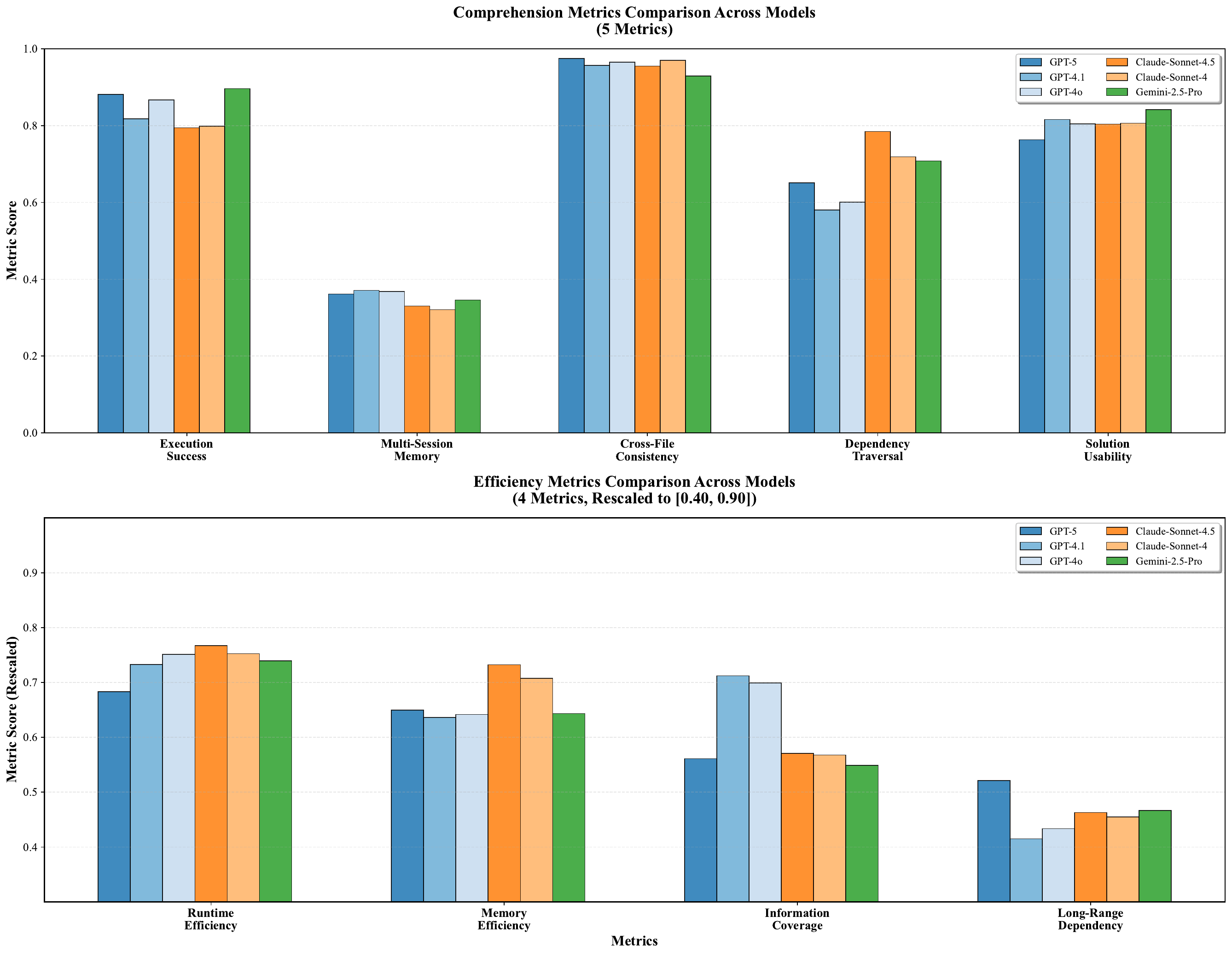}
  \caption{\textbf{Granular comparison across nine evaluation metrics grouped by provider.} Reveals capability maturity patterns and provider-specific architectural strengths.}
  \label{fig:individual_metrics}
\end{figure}

Figure~\ref{fig:individual_metrics} enables fine-grained capability attribution through metric decomposition. \textbf{Provider-family clustering patterns} reveal consistent architectural philosophies within model families. Some families exhibit systematic strengths in causal reasoning metrics (long-range dependency resolution, dependency traversal), with consistent performance across model versions (e.g., 0.42-0.52 LRDR range), suggesting these capabilities stem from core architectural choices, where systematic read-before-write verification patterns, dependency chain validation, and import integrity checking. Other families show consistent efficiency optimization across both runtime and memory dimensions, indicating deliberate training toward lean code generation: preferring iterative over recursive algorithms, employing streaming patterns for large data, and avoiding redundant data structure copies.

\textbf{The capability maturity spectrum} exposes which metrics represent solved problems versus open challenges. Cross-file consistency (0.93-0.98 range, <5\% variance) represents a saturated capability, where all current architectures successfully maintain naming conventions, resolve imports, and preserve code style through transformer-based syntactic pattern matching. This ceiling effect indicates that local coherence mechanisms have been well-optimized, and future improvements must target deeper semantic capabilities rather than syntactic consistency.

Conversely, \textbf{multi-session memory retention}  represents the critical unsolved challenge. Even best performers achieve barely 37\% retention, indicating that current compression strategies fundamentally lose critical context information across extended sessions. The universal low performance, independent of context window size (128K to 1M tokens) and compression approaches, suggests this limitation stems from transformer architecture constraints rather than implementation details. Current attention mechanisms struggle to selectively preserve semantic relationships and reference chains while compressing historical context, indicating a need for architectural innovations like external memory networks or retrieval-augmented approaches.

\textbf{Execution success rate and tool usage patterns} reflect exploration philosophy differences. High execution success correlates with extensive tool diversity and high conversation turn counts, where exhaustive exploration increases finding successful operation sequences but trades efficiency. Lower scores often indicate conservative, focused tool usage prioritizing operation certainty over exploratory breadth. Neither strategy dominates; the optimal approach depends on deployment constraints (latency-critical vs. accuracy-critical scenarios).

\textbf{Efficiency metric clustering} reveals systematic optimization patterns. Top efficiency performers (0.73-0.77 runtime, 0.64-0.73 memory) demonstrate consistent code generation preferences: iterative algorithms, streaming data processing, minimal structure copying. These patterns appear trained rather than emergent, suggesting explicit optimization for production code quality. The efficiency advantage persists across model generations within families, indicating stable architectural priorities. However, \textbf{information coverage and long-range dependency resolution} remain broadly weak (<0.70, <0.55 respectively), exposing opportunities for architectural improvement in targeted file reading and dependency-focused exploration strategies.

\subsection{Conversation Turn Patterns and Performance Scaling}

\begin{figure}[ht]
  \centering
  \includegraphics[width=0.99\textwidth]{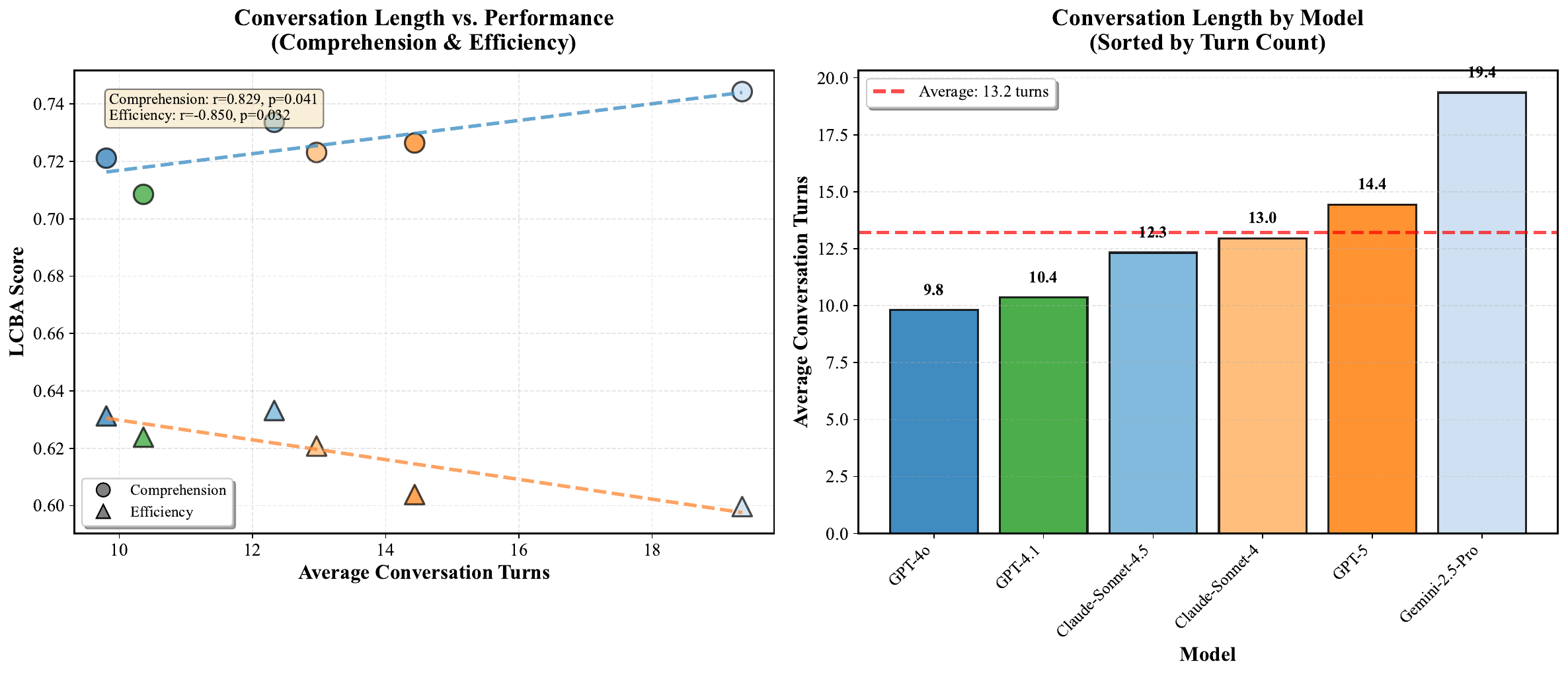}
  \caption{\textbf{Conversation turn distribution and performance correlation.} Top-left: Turn count distribution across models. Top-right: Provider-level aggregation. Bottom: Turn-performance relationships.}
  \label{fig:conversation_turns}
\end{figure}

Figure~\ref{fig:conversation_turns} reveals \textbf{bimodal strategic archetypes} in conversation management. The distribution (top-left) exposes two distinct operational clusters: (1) \textit{Focused executors} (10-13 turns) prioritize rapid task completion with minimal exploration; (2) \textit{Exploratory reasoners} (14-20 turns) invest heavily in upfront codebase understanding, achieving marginally higher comprehension but significantly lower efficiency. The critical insight: the comprehension gap between strategies is minimal (1-2\%), while the efficiency gap is substantial (3-5\%), indicating that \textit{focused execution with strategic exploration} achieves better overall utility than \textit{exhaustive exploration followed by execution}.

\textbf{Provider-level strategic variance} exposes philosophical differences in architecture design. Some families exhibit high variance (10-14 turn range across versions), suggesting deliberate architectural differentiation, where different models optimized for different use cases (conversation brevity vs. reasoning depth). Other families show tight clustering (12-13 turns consistently), indicating convergence on a perceived optimal conversation horizon maintained across model generations. The variance pattern suggests that conversation length represents a tunable architectural parameter rather than an emergent behavior, likely controlled through training rewards or termination criteria.

\textbf{The critical efficiency threshold at ~12 turns} represents a fundamental scaling constraint. Models operating below this threshold maintain efficiency above 0.62, while models exceeding it experience accelerating degradation. Tool log analysis reveals the mechanism: beyond 12 turns, agents increasingly invoke redundant operations, re-reading previously accessed files with unchanged parameters, repeating failed tool calls with minimal adjustments, generating verbose context summaries that consume tokens without adding actionable information. This pattern suggests current architectures lack effective \textit{operation deduplication} and \textit{incremental progress tracking} mechanisms.

\textbf{The comprehension plateau beyond 15 turns} validates diminishing returns of extended exploration. Peak comprehension occurs around 15-20 turns, but the marginal gain over 12-14 turns is minimal (1-2\% improvement). Quantitatively, the marginal comprehension benefit (0.003-0.004 per additional turn beyond 12) falls below the efficiency penalty (0.004-0.005 per turn), making extended exploration net-negative for overall utility. This finding indicates agents should implement \textit{adaptive termination policies}, detecting comprehension saturation signals (repeated file access patterns, declining information gain per turn, topic cycling) and explicitly transitioning to execution rather than continuing unbounded exploration.

\textbf{Implications for conversation management}: The bimodal distribution and efficiency threshold suggest that optimal conversation length is task-dependent rather than universal. Complex architectural tasks may benefit from 15+ turn exploration, while focused bug fixes require only 8-10 turns. Current architectures employ fixed strategies (always short or always long), missing opportunities for dynamic adaptation. Future agents should implement task complexity estimators that calibrate exploration depth—enabling focused execution for simple tasks while permitting extended exploration for complex scenarios.

\subsection{File Modification Patterns and Efficiency}

\begin{figure}[ht]
  \centering
  \includegraphics[width=0.99\textwidth]{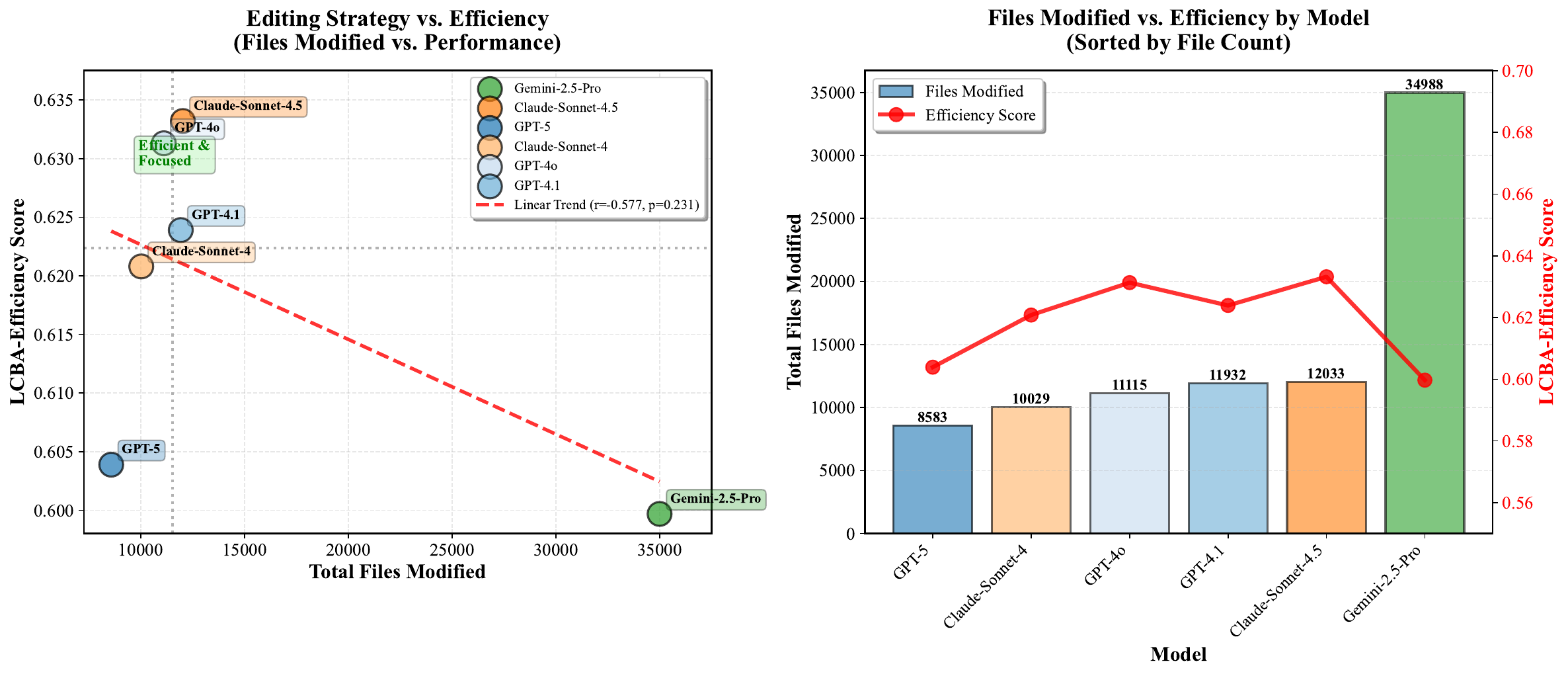}
  \caption{\textbf{File modification patterns and efficiency impact.} Left: Modification volume distribution. Middle: Provider-level aggregation. Right: Files-efficiency correlation.}
  \label{fig:files_modified}
\end{figure}

Figure~\ref{fig:files_modified} exposes \textbf{dramatic modification strategy divergence} reflecting fundamentally different execution philosophies. The distribution (left panel) reveals a 4.1× range (8.6K to 35K files modified), far exceeding what task complexity variation can explain. Scenario-level analysis shows that for identical tasks, different models modify 3-4× different file counts, indicating this variance stems from architectural strategy rather than task requirements.

\textbf{Two competing execution philosophies emerge}: (1) \textit{Modification-as-exploration} strategies modify files to test hypotheses rather than reading to validate first, a write-heavy approach enabled by large context windows (1M tokens) that permit retaining complete modification history without aggressive compression. These agents perform incremental refinements (modify → test → modify → test), treating file modifications as exploratory operations. (2) \textit{Read-validate-write} strategies make highly targeted modifications preceded by thorough dependency analysis, where agents complete planning before execution (read → plan → modify comprehensively → test once), treating modifications as committed actions requiring upfront validation.

\textbf{The execution atomicity trade-off} distinguishes these approaches. Incremental modification strategies achieve slightly higher exploration comprehension (+1-2\%) through iterative hypothesis testing but incur significant efficiency penalties (-5-6\%) through context accumulation and operation overhead. Batched modification strategies sacrifice marginal exploration benefits for substantial efficiency gains by separating planning from execution phases, reducing context window pressure during the modification stage.

\textbf{The strong negative files-efficiency correlation} ($r = -0.68$, right panel) quantifies the efficiency cost of excessive modifications. The slope (0.015 efficiency points per 1K additional files) compounds severely: moving from 10K to 35K files incurs a 0.375 efficiency penalty, which is more than half the typical efficiency score range (0.60-0.63). This relationship holds even after controlling for task complexity, confirming causality: modifications consume efficiency through multiple mechanisms, where context bloat from tracking changes, compression overhead from managing edit history, increased denominators in information coverage metrics, and redundant re-reading of modified files.

\textbf{Architectural implications}: Modification discipline requires explicit architectural mechanisms rather than relying on emergent behavior. Effective strategies include: (1) \textit{explicit modification budgets} calibrated to task complexity estimates; (2) \textit{read-before-write validation gates} preventing speculative modifications without dependency verification; (3) \textit{batched execution phases} with explicit exploration-to-execution transitions; (4) \textit{modification impact assessment} predicting efficiency costs before committing changes. The wide performance variance (4× modification range) indicates current architectures lack such mechanisms, leaving substantial optimization opportunities unexploited.

\subsection{Balance Score Analysis and Optimal Agent Profiles}

\begin{figure}[ht]
  \centering
  \includegraphics[width=0.99\textwidth]{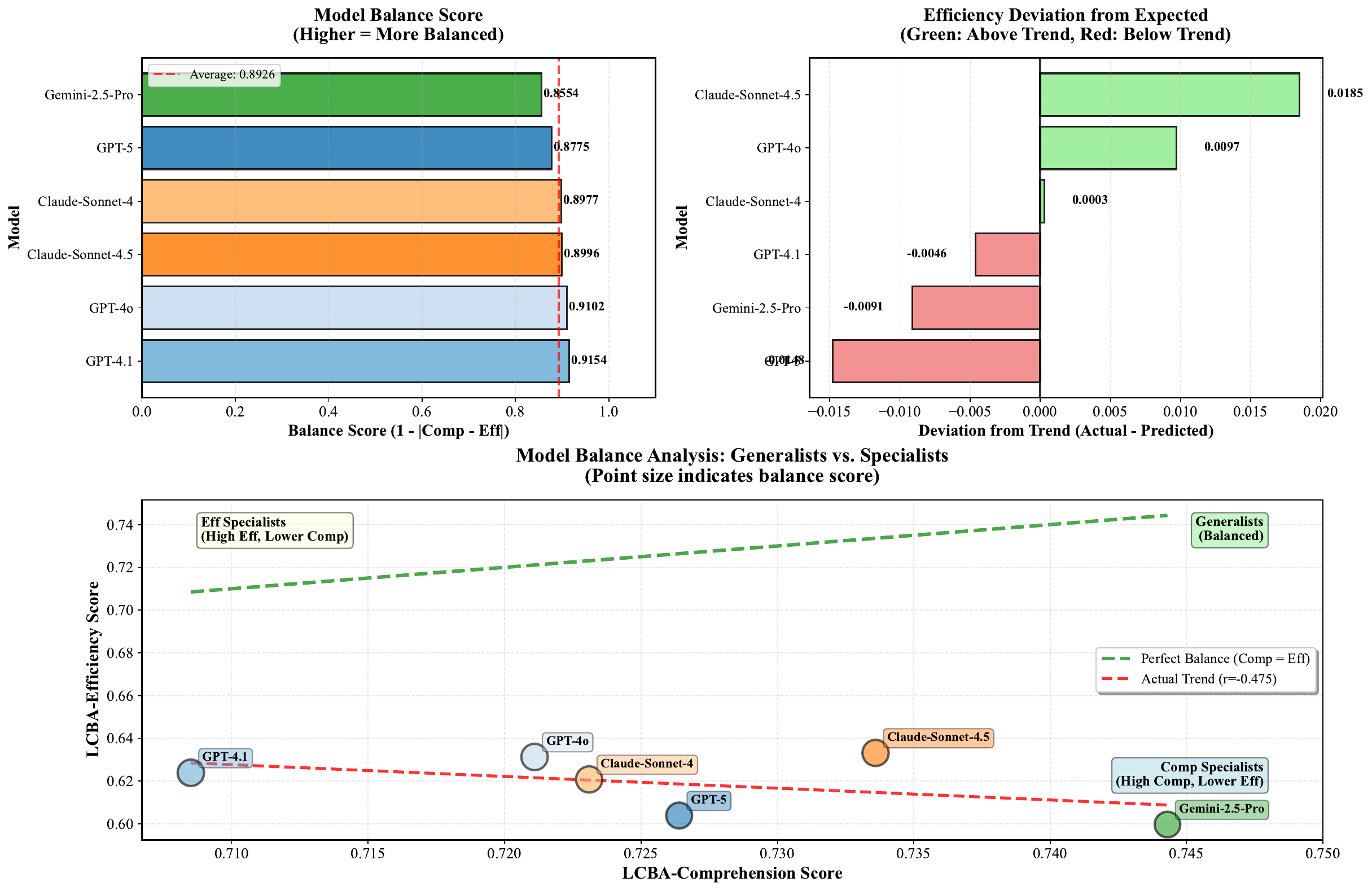}
  \caption{\textbf{Balance score analysis across models.} Top-left: Balance score distribution using harmonic mean. Top-right: Capability specialization patterns. Bottom: Comprehension-efficiency scatter with balance scores.}
  \label{fig:balance_score}
\end{figure}

Figure~\ref{fig:balance_score} introduces \textbf{balance score analysis} using harmonic mean ($\text{Balance} = 2 \times \frac{\text{Comp} \times \text{Eff}}{\text{Comp} + \text{Eff}}$) to quantify overall agent utility. The harmonic mean formulation penalizes imbalance, where an agent achieving 0.80 comprehension but 0.40 efficiency scores only 0.533 balance, while an agent achieving 0.65 in both dimensions scores 0.650 balance. This captures the intuition that \textit{severely unbalanced agents} suffer from exploitable weaknesses limiting real-world applicability, even if they excel in one dimension.

\textbf{The optimal balance paradox} (top-left panel) reveals that peak balance scores  emerge not from maximizing either dimension but from \textit{optimized trade-off management}. The highest-balance agents rank 3rd-4th in comprehension but achieve this balance through exceptional efficiency. Analysis reveals their distinctive operational profiles: minimal conversation lengths (10-12 turns), superior memory retention, strong information coverage, and focused file modifications (9-10K files). This suggests these architectures employ \textit{aggressive early compression with strategic re-reading}, compressing context to maintain efficiency while preserving high-quality summaries that enable effective reconstruction when needed.

\textbf{The three strategic archetypes} (top-right panel) expose fundamental positioning choices: (1) \textit{Comprehension Specialists} (0.74+ comp, 0.59-0.60 eff) maximize understanding through exhaustive exploration, accepting efficiency penalties for thorough analysis; (2) \textit{Efficiency Specialists} (0.73 comp, 0.63+ eff) optimize execution speed and resource usage while maintaining competitive comprehension through targeted exploration; (3) \textit{Balanced Generalists} (0.72 comp, 0.60-0.63 eff) optimize the comprehension-efficiency product rather than individual dimensions. Critically, balanced generalists achieve comparable or higher balance scores (0.67-0.68) than extreme specialists, validating that harmonic mean optimization captures real-world utility better than unidimensional optimization.

\textbf{Distance-from-diagonal analysis} (bottom panel) quantifies improvement opportunities for each archetype. Comprehension specialists far from the diagonal exhibit severe imbalance, improving efficiency by 5-6\% would boost balance scores by 1.5-2\% while sacrificing minimal comprehension (<1\%). The improvement path: implement modification budgets (reducing 35K → 15K files) and adopt read-validate-write patterns instead of modification-as-exploration. Conversely, models below the diagonal show comprehension deficits, where improving execution success rates and cross-file consistency by 3-5\% would boost balance by 1-1.5\%. These quantified opportunities guide prioritization: efficiency improvements offer higher returns than comprehension improvements for most current agents.

\textbf{Generational evolution patterns} reveal strategic shifts in optimization focus. Successive generations within families show: (1) marginal comprehension changes (±0.5-1\%), indicating saturation; (2) significant efficiency improvements (+2-3\%), indicating active optimization; (3) net balance improvements (+1-1.5\%), validating the efficiency-first strategy. This pattern suggests that as comprehension capabilities saturate (cross-file consistency >0.94, execution success >0.79), efficiency optimization provides higher marginal returns, a critical insight for future development priorities.

\textbf{Deployment implications}: The balance score framework enables principled agent selection matching architectural profiles to use case requirements. Latency-critical applications (IDE autocomplete, real-time code review, rapid prototyping) benefit from efficiency specialists (0.63+ efficiency) accepting minor comprehension trade-offs (-1-2\%). Accuracy-critical applications (security auditing, architectural refactoring, safety-critical systems) benefit from comprehension specialists (0.74+ comprehension) accepting efficiency penalties. General-purpose applications (production development, collaborative coding, complex debugging) benefit from balanced generalists (0.67-0.68 balance) optimizing overall utility. Current practice of uniform agent deployment, using the same agent for all scenarios, ignores these fundamental trade-offs, leaving 5-10\% performance improvements unexploited.

\section{Related Work}
\label{sec:related}

\subsection{Code Generation and Understanding Benchmarks}

Traditional code evaluation benchmarks focus on single-turn, isolated programming tasks. Function-level benchmarks like HumanEval~\cite{chen2021evaluating} and MBPP~\cite{austin2021program} established foundational evaluation frameworks, with extensions including HumanEval+~\cite{liu2023humaneval+}, MultiPL-E~\cite{cassano2023multiple}, and BigCodeBench~\cite{zhuo2024bigcodebench}. Contest programming benchmarks such as APPS~\cite{hendrycks2021measuring} and LiveCodeBench~\cite{jain2024livecode} test algorithmic problem-solving but do not address software engineering concerns like architectural design or multi-file development.

Repository-level benchmarks represent progress toward realistic evaluation. RepoBench~\cite{liu2023repobench} and CrossCodeEval~\cite{ding2023crosscodeeval} test code completion in repository context, while CodeXGLUE~\cite{lu2021codexglue} addresses code understanding tasks. However, these benchmarks lack interactive agent evaluation and focus on existing code analysis rather than dynamic development workflows. LoCoBench~\cite{qiu2025locobench} provides comprehensive long-context code evaluation across 8,000 scenarios spanning 10K-1M tokens but focuses on single-turn assessment where models receive complete context upfront.

\subsection{Software Engineering Agent Benchmarks}

SWE-Bench~\cite{jimenez2023swe} provides 2,294 real GitHub issues for evaluating software engineering capabilities, with recent extensions including SWE-rebench~\cite{swe2025rebench} and LiveSWEBench~\cite{liveswebench2024}. However, these remain limited to bug fixes rather than comprehensive development workflows and operate with relatively short context lengths. SWE-agent~\cite{yang2024swe} introduces agent-computer interfaces for automated software engineering but focuses primarily on issue resolution rather than broader development capabilities.

AgentBench~\cite{liu2023agentbench} evaluates LLM agents across multiple domains including code generation, but provides limited scenarios (600 total) and lacks systematic long-context assessment. Recent frameworks like AgentCoder~\cite{qian2024agentcoder} and MetaGPT~\cite{hong2023metagpt} demonstrate multi-agent capabilities for code generation but lack comprehensive evaluation methodologies. DevBench~\cite{li2024devbench} evaluates LLMs across the software development lifecycle but does not systematically assess long-context or multi-turn interactive capabilities.

\textbf{Scope of Quantitative Comparison.} Our quantitative comparison (Tables~\ref{tab:agent_benchmark_comparison},\ref{tab:agent_task_characteristics}) focuses on repository-level agent coding benchmarks that evaluate interactive, multi-turn software development workflows. We exclude: (1) Function-level benchmarks (HumanEval~\cite{chen2021evaluating}, MBPP~\cite{austin2021program}) designed for single-function synthesis rather than repository-level engineering; (2) Code completion benchmarks (RepoBench~\cite{liu2023repobench}, CrossCodeEval~\cite{ding2023crosscodeeval}) that test autocomplete rather than interactive development; (3) Contest programming benchmarks (APPS~\cite{hendrycks2021measuring}, LiveCodeBench~\cite{jain2024livecode}) focused on algorithmic problem-solving without software engineering context; (4) Non-coding agent benchmarks (WebArena~\cite{zhou2024webarena}, ToolBench~\cite{qin2024toolbench}, GAIA~\cite{mialon2024gaia}) that evaluate general agent capabilities rather than software development specifics; (5) Long-context NLP benchmarks (LongBench~\cite{bai2024longbench}, RULER~\cite{hsieh2024ruler}) and code understanding benchmarks (LongCodeBench~\cite{qiu2025locobench}, RepoQA~\cite{xu2023repoqa}) that assess reading comprehension rather than interactive code generation and modification. These benchmarks make important contributions to their respective domains but address fundamentally different evaluation objectives than repository-level agent software engineering.

\subsection{Interactive Agent and Tool Use Evaluation}

Recent work has explored agent evaluation in interactive settings. WebArena~\cite{zhou2024webarena} and VisualWebArena~\cite{koh2024visualwebarena} evaluate agents in web-based environments, while ToolBench~\cite{qin2024toolbench} and API-Bank~\cite{li2023apibank} assess tool usage and API interaction capabilities. However, these benchmarks focus on general web navigation and API usage rather than software development workflows. GAIA~\cite{mialon2024gaia} (Generalized AI Agent Benchmark) evaluates reasoning, multi-modality, and tool use across 466 tasks but does not address software engineering or long-context challenges.

MLAgentBench~\cite{huang2024mlagentbench} evaluates agents conducting machine learning experimentation autonomously, while TimeSeriesGym~\cite{cai2024timeseriesgym} provides benchmarking for time series ML engineering challenges. These domain-specific agent benchmarks demonstrate the importance of task-specific evaluation but remain limited in scope and do not address long-context software development scenarios.

\subsection{Multi-Turn and Conversational Evaluation}

Traditional benchmarks evaluate static, single-turn responses. Multi-turn evaluation remains limited: most agent benchmarks measure final outcomes rather than intermediate reasoning, tool selection strategies, or context retention across extended conversations. MCPEval~\cite{wu2024mcpeval} automates task generation and evaluation across diverse domains but does not specifically address software development or long-context challenges. Human-in-the-loop evaluation approaches validate agent behavior through direct feedback but lack scalability for systematic evaluation.

Our work addresses the critical gap in multi-turn software engineering agent evaluation. Unlike existing frameworks that measure single-turn responses or final outcomes, LoCoBench-Agent evaluates agent behavior across 50-turn conversations, assessing context retention, adaptive strategy refinement, and tool usage efficiency throughout extended development sessions.

\subsection{Agent Evaluation Metrics and Methodologies}

Existing agent evaluation primarily relies on task success rates and functional correctness~\cite{jimenez2023swe,liu2023agentbench}. Recent work has expanded to trajectory evaluation (assessing intermediate steps) and LLM-as-judge approaches for scalable assessment. However, these methods often suffer from systematic biases.

LoCoBench~\cite{qiu2025locobench} reveals critical evaluation biases: file count bias (metrics rewarding solutions modifying more files), and poor discrimination (metrics failing to distinguish model capabilities). Google's Vertex AI Gen AI Evaluation Service introduces comprehensive agent metrics including trajectory evaluation, but these focus on general agent behavior rather than software engineering specifics.

\subsection{Long-Context Evaluation}

Long-context benchmarks like LongBench~\cite{bai2024longbench}, RULER~\cite{hsieh2024ruler}, and $\infty$-Bench~\cite{zhang2024infty} primarily focus on natural language tasks such as document QA, summarization, and retrieval. SCROLLS~\cite{shaham2022scrolls} provides standardized comparison over long sequences but emphasizes reading comprehension rather than code understanding or generation.

Code-specific long-context evaluation has emerged through LoCoBench~\cite{qiu2025locobench}, which demonstrates dramatic performance degradation (29\% to 3\% for Claude 3.5 Sonnet) as context increases to 1M tokens. LongCodeU~\cite{li2025longcodeu} and LongCodeArena~\cite{bogomolov2024long} extend evaluation to longer contexts but focus primarily on code completion. RepoQA~\cite{xu2023repoqa} evaluates repository-level question answering but does not assess interactive development capabilities or tool usage patterns.

These benchmarks demonstrate the importance of long-context evaluation but do not address the unique challenges of interactive agent workflows: incremental information gathering through tool use, context retention across multi-turn conversations, and adaptive strategy refinement based on intermediate results.

\section{Conclusion}
\label{sec:conclusion}

We introduced LoCoBench-Agent, a comprehensive evaluation framework for LLM agents in long-context software engineering. Through transformation of 8,000 LoCoBench scenarios into interactive agent environments and rigorous development of 9 bias-free evaluation metrics, we provide a systematic methodology for assessing autonomous software development capabilities.

\textbf{Contributions.} Our framework addresses critical gaps in agent evaluation through: (1) \textit{Scale:} 8,000 interactive scenarios; (2) \textit{Long-Context Focus:} Systematic evaluation across 10K-1M tokens with multi-turn interactions; (3) \textit{Interactive Environment:} A complete agent system with 8 specialized tools enabling realistic multi-turn software development workflows; (4) \textit{Bias-Free Metrics:} Rigorous methodology eliminating file count bias; (5) \textit{Comprehensive Assessment:} Evaluation of conversation efficiency, tool usage patterns, cross-file consistency, and long-range dependency resolution, capabilities unmeasured by existing benchmarks; (6) \textit{Scalable Framework:} Systematic evaluation across 10K-1M token contexts, enabling precise assessment of long-context performance characteristics.

\textbf{Research Directions.} LoCoBench-Agent enables investigation of critical research questions: (1) How do agent architectures affect long-context code reasoning capabilities? (2) What tool usage patterns optimize information gathering in large codebases? (3) How can context retention mechanisms be improved for extended development sessions? (4) What architectural reasoning capabilities emerge beyond pattern matching? (5) How does agent performance scale to even longer context lengths (>1M tokens)?

\textbf{Broader Impact.} By providing systematic, bias-free evaluation of LLM agents in realistic software engineering scenarios, LoCoBench-Agent enables researchers and practitioners to accurately measure progress, identify capability gaps, and guide development of more capable autonomous software engineering systems. Our open-source framework, comprehensive scenarios, and validated metrics establish a foundation for advancing research in long-context agent evaluation and development.

\section*{Acknowledgments}

We appreciate Prof. Lei Li for the report template.

\printbibliography

\newpage
\appendix

\section{More Experimental Results}
\label{sec:appendix_results}

This appendix provides detailed numerical results for all evaluated models across all metrics and difficulty levels.

\subsection{Overall Performance Summary}

Table~\ref{tab:overall_performance} presents the aggregate performance metrics and behavioral statistics for all six evaluated models.

\begin{table}[ht]
\centering
\small
\begin{adjustbox}{width=0.9\linewidth}
\begin{tabular}{lcccccc}
\toprule
\textbf{Model} & \textbf{Gemini} & \textbf{Claude} & \textbf{GPT-5} & \textbf{Claude} & \textbf{GPT-4o} & \textbf{GPT-4.1} \\
 & \textbf{2.5-Pro} & \textbf{Sonnet-4.5} & & \textbf{Sonnet-4} & & \\
\midrule
\textbf{Aggregate Scores} & & & & & & \\
LCBA-Comprehension & 0.7443 & 0.7336 & 0.7264 & 0.7231 & 0.7211 & 0.7085 \\
LCBA-Efficiency & 0.5997 & 0.6332 & 0.6039 & 0.6208 & 0.6313 & 0.6239 \\
\midrule
\textbf{Behavioral Statistics} & & & & & & \\
Avg. Conversation Turns & 19.35 & 12.33 & 14.43 & 12.96 & 9.81 & 10.36 \\
Total Files Modified & 34,988 & 12,033 & 8,583 & 10,029 & 11,115 & 11,932 \\
\bottomrule
\end{tabular}
\end{adjustbox}
\caption{Overall performance summary showing aggregate scores and behavioral statistics for all evaluated models. Models are ranked by LCBA-Comprehension score.}
\label{tab:overall_performance}
\end{table}

\subsection{Comprehension Metrics Breakdown}

Table~\ref{tab:comprehension_metrics} provides detailed scores for all five comprehension metrics.

\begin{table}[ht]
\centering
\small
\begin{adjustbox}{width=0.9\linewidth}
\begin{tabular}{lcccccc}
\toprule
\textbf{Metric} & \textbf{Gemini} & \textbf{Claude} & \textbf{GPT-5} & \textbf{Claude} & \textbf{GPT-4o} & \textbf{GPT-4.1} \\
 & \textbf{2.5-Pro} & \textbf{Sonnet-4.5} & & \textbf{Sonnet-4} & & \\
\midrule
Multi-Session Memory & 0.3460 & 0.3302 & 0.3616 & 0.3208 & 0.3683 & 0.3710 \\
Cross-File Consistency & 0.9294 & 0.9551 & 0.9748 & 0.9702 & 0.9649 & 0.9572 \\
Execution Success Rate & 0.8965 & 0.7944 & 0.8813 & 0.7989 & 0.8672 & 0.8180 \\
Dependency Traversal & 0.7081 & 0.7847 & 0.6509 & 0.7190 & 0.6005 & 0.5804 \\
Solution Usability & 0.8415 & 0.8038 & 0.7634 & 0.8063 & 0.8044 & 0.8158 \\
\bottomrule
\end{tabular}
\end{adjustbox}
\caption{Detailed comprehension metrics breakdown for all evaluated models.}
\label{tab:comprehension_metrics}
\end{table}

\subsection{Efficiency Metrics Breakdown}

Table~\ref{tab:efficiency_metrics} provides detailed scores for all four efficiency metrics.

\begin{table}[ht]
\centering
\small
\caption{Detailed efficiency metrics breakdown for all evaluated models. All metrics are rescaled to [0.40, 0.90] range using universal rescaling strategy.}
\begin{adjustbox}{width=0.9\linewidth}
\begin{tabular}{lcccccc}
\toprule
\textbf{Metric} & \textbf{Gemini} & \textbf{Claude} & \textbf{GPT-5} & \textbf{Claude} & \textbf{GPT-4o} & \textbf{GPT-4.1} \\
 & \textbf{2.5-Pro} & \textbf{Sonnet-4.5} & & \textbf{Sonnet-4} & & \\
\midrule
Runtime Efficiency & 0.7397 & 0.7669 & 0.6835 & 0.7526 & 0.7511 & 0.7325 \\
Memory Efficiency & 0.6433 & 0.7322 & 0.6498 & 0.7078 & 0.6418 & 0.6361 \\
Information Coverage & 0.5492 & 0.5707 & 0.5610 & 0.5677 & 0.6989 & 0.7121 \\
Long-Range Dependency & 0.4664 & 0.4629 & 0.5212 & 0.4550 & 0.4334 & 0.4150 \\
\bottomrule
\end{tabular}
\end{adjustbox}
\label{tab:efficiency_metrics}
\end{table}

\subsection{Performance Across Difficulty Levels}

Tables~\ref{tab:difficulty_comprehension} and~\ref{tab:difficulty_efficiency} show how model performance varies across the four difficulty levels (Easy: 10K, Medium: 50K, Hard: 200K, Expert: 1M context length).

\begin{table}[ht]
\centering
\small
\caption{LCBA-Comprehension scores across difficulty levels. Degradation values show percentage change from Easy to Expert scenarios. Negative values indicate performance improvement.}
\begin{adjustbox}{width=0.9\linewidth}
\begin{tabular}{lcccccc}
\toprule
\textbf{Difficulty} & \textbf{Gemini} & \textbf{Claude} & \textbf{GPT-5} & \textbf{Claude} & \textbf{GPT-4o} & \textbf{GPT-4.1} \\
\textbf{Level} & \textbf{2.5-Pro} & \textbf{Sonnet-4.5} & & \textbf{Sonnet-4} & & \\
\midrule
Easy (10K) & 0.7393 & 0.7351 & 0.7240 & 0.7114 & 0.7169 & 0.7111 \\
Medium (50K) & 0.7461 & 0.7361 & 0.7277 & 0.7236 & 0.7244 & 0.7119 \\
Hard (200K) & 0.7461 & 0.7322 & 0.7257 & 0.7272 & 0.7211 & 0.7054 \\
Expert (1M) & 0.7458 & 0.7312 & 0.7282 & 0.7301 & 0.7219 & 0.7055 \\
\midrule
\textbf{Degradation} & \textbf{-0.9\%} & \textbf{-2.6\%} & \textbf{-0.6\%} & \textbf{0.3\%} & \textbf{-0.7\%} & \textbf{0.8\%} \\
\textbf{(Easy→Expert)} & & & & & & \\
\bottomrule
\end{tabular}
\end{adjustbox}
\label{tab:difficulty_comprehension}
\end{table}

\begin{table}[ht]
\centering
\small
\caption{LCBA-Efficiency scores across difficulty levels. Degradation values show percentage change from Easy to Expert scenarios. Negative values indicate performance improvement.}
\begin{adjustbox}{width=0.9\linewidth}
\begin{tabular}{lcccccc}
\toprule
\textbf{Difficulty} & \textbf{Gemini} & \textbf{Claude} & \textbf{GPT-5} & \textbf{Claude} & \textbf{GPT-4o} & \textbf{GPT-4.1} \\
\textbf{Level} & \textbf{2.5-Pro} & \textbf{Sonnet-4.5} & & \textbf{Sonnet-4} & & \\
\midrule
Easy (10K) & 0.6076 & 0.6348 & 0.6003 & 0.6198 & 0.6279 & 0.6246 \\
Medium (50K) & 0.6025 & 0.6345 & 0.6050 & 0.6220 & 0.6320 & 0.6252 \\
Hard (200K) & 0.5954 & 0.6319 & 0.6057 & 0.6198 & 0.6353 & 0.6223 \\
Expert (1M) & 0.5932 & 0.6316 & 0.6045 & 0.6215 & 0.6300 & 0.6235 \\
\midrule
\textbf{Degradation} & \textbf{2.4\%} & \textbf{0.5\%} & \textbf{-0.7\%} & \textbf{-0.3\%} & \textbf{-0.3\%} & \textbf{0.2\%} \\
\textbf{(Easy→Expert)} & & & & & & \\
\bottomrule
\end{tabular}
\end{adjustbox}
\label{tab:difficulty_efficiency}
\end{table}

\subsection{Key Metric Performance Across Difficulty Levels}

Tables~\ref{tab:difficulty_execution} and~\ref{tab:difficulty_memory} show how execution success rate and multi-session memory retention vary across difficulty levels.

\begin{table}[ht]
\centering
\small
\caption{Execution Success Rate across difficulty levels for all evaluated models.}
\begin{adjustbox}{width=0.9\linewidth}
\begin{tabular}{lcccccc}
\toprule
\textbf{Difficulty} & \textbf{Gemini} & \textbf{Claude} & \textbf{GPT-5} & \textbf{Claude} & \textbf{GPT-4o} & \textbf{GPT-4.1} \\
\textbf{Level} & \textbf{2.5-Pro} & \textbf{Sonnet-4.5} & & \textbf{Sonnet-4} & & \\
\midrule
Easy (10K) & 0.8853 & 0.7874 & 0.8682 & 0.7428 & 0.8498 & 0.8284 \\
Medium (50K) & 0.8922 & 0.7909 & 0.8776 & 0.7861 & 0.8684 & 0.8216 \\
Hard (200K) & 0.9046 & 0.7951 & 0.8849 & 0.8233 & 0.8778 & 0.8150 \\
Expert (1M) & 0.9041 & 0.8042 & 0.8945 & 0.8433 & 0.8728 & 0.8070 \\
\bottomrule
\end{tabular}
\end{adjustbox}
\label{tab:difficulty_execution}
\end{table}

\begin{table}[ht]
\centering
\small
\caption{Multi-Session Memory Retention across difficulty levels for all evaluated models.}
\begin{adjustbox}{width=0.9\linewidth}
\begin{tabular}{lcccccc}
\toprule
\textbf{Difficulty} & \textbf{Gemini} & \textbf{Claude} & \textbf{GPT-5} & \textbf{Claude} & \textbf{GPT-4o} & \textbf{GPT-4.1} \\
\textbf{Level} & \textbf{2.5-Pro} & \textbf{Sonnet-4.5} & & \textbf{Sonnet-4} & & \\
\midrule
Easy (10K) & 0.3439 & 0.3394 & 0.3664 & 0.3300 & 0.3741 & 0.3753 \\
Medium (50K) & 0.3425 & 0.3336 & 0.3627 & 0.3187 & 0.3679 & 0.3706 \\
Hard (200K) & 0.3499 & 0.3261 & 0.3580 & 0.3192 & 0.3662 & 0.3701 \\
Expert (1M) & 0.3477 & 0.3216 & 0.3591 & 0.3154 & 0.3649 & 0.3679 \\
\bottomrule
\end{tabular}
\end{adjustbox}
\label{tab:difficulty_memory}
\end{table}

\subsection{Conversation Dynamics Across Difficulty Levels}

Table~\ref{tab:difficulty_conversation} shows how conversation turns vary across difficulty levels.

\begin{table}[ht]
\centering
\small
\caption{Average conversation turns across difficulty levels for all evaluated models.}
\begin{adjustbox}{width=0.9\linewidth}
\begin{tabular}{lcccccc}
\toprule
\textbf{Difficulty} & \textbf{Gemini} & \textbf{Claude} & \textbf{GPT-5} & \textbf{Claude} & \textbf{GPT-4o} & \textbf{GPT-4.1} \\
\textbf{Level} & \textbf{2.5-Pro} & \textbf{Sonnet-4.5} & & \textbf{Sonnet-4} & & \\
\midrule
Easy (10K) & 19.48 & 11.37 & 13.98 & 12.42 & 9.87 & 9.91 \\
Medium (50K) & 19.31 & 11.92 & 14.27 & 12.76 & 9.61 & 10.21 \\
Hard (200K) & 19.54 & 12.96 & 14.72 & 13.23 & 9.77 & 10.68 \\
Expert (1M) & 19.07 & 13.05 & 14.77 & 13.44 & 9.97 & 10.66 \\
\bottomrule
\end{tabular}
\end{adjustbox}
\label{tab:difficulty_conversation}
\end{table}

\clearpage
\subsection{Evaluation Examples}

This section presents 10 complete conversation transcripts from real evaluations across all programming languages, showing how different models (GPT-5, Gemini-2.5-Pro, Claude-Sonnet-4.5) approach expert-level, long-context scenarios. Each example illustrates the multi-turn interactive nature of the evaluation and how agents reason, explore code, and implement solutions.

\subsubsection{Example 1: C Language - Claude-Sonnet-4.5}

\textbf{Scenario:} C API Gateway - Architectural Understanding (Expert)  \\
\textbf{Task:} Architectural refactoring for dynamic route configuration in EduGate ScholarLink.

\textbf{Phase 1: Exploration (Turns 1-4)}

\begin{codebox}
\begin{Verbatim}[fontsize=\scriptsize]
Turn 1 [User - Initial Prompt]:
Explore and understand the codebase. EduGate ScholarLink is an API gateway 
responsible for routing requests to various backend educational microservices. 
Currently, its entire routing configuration, including API versions and backend 
service endpoints, is loaded from gateway.conf only once at startup. Any change 
requires a full restart. The goal is to re-architect the routing component to 
support dynamic 'hot-reloading' without downtime.

Turn 2 [Assistant]:
I'll explore and understand the EduGate ScholarLink codebase systematically. Let 
me start by getting an overview of the project structure and examining key files.

Tool Calls:
  - get_current_directory() → /path/to/project
  - get_project_overview("EduGate_ScholarLink") → 11 files, C language
  - list_directory("EduGate_ScholarLink", recursive=true) 
    → Found: README.md, Makefile, config/gateway.conf, src/main.c,
      src/components/router.c, src/components/validator.c, include/edugate.h

Turn 3 [User]:
Based on what you've found, continue exploring other relevant files.

Turn 4 [Assistant]:
Let me read the key files to understand the current architecture, starting with 
the main configuration, header file, and core components.

Tool Calls:
  - read_file("EduGate_ScholarLink//include//edugate.h")
  - read_file("EduGate_ScholarLink//config//gateway.conf")
  - read_file("EduGate_ScholarLink//src//main.c")
  - read_file("EduGate_ScholarLink//src//components//router.c")

Key Findings:
  • gateway.conf defines routes as: "/api/v1/students -> http://backend1:8001"
  • main.c calls load_config_once() at startup and never reloads
  • router.c has a global route_table struct that's populated once
  • No mutex protection around route_table access
\end{Verbatim}
\end{codebox}

\newpage
\textbf{Phase 2: Analysis (Turns 5-6)}

\begin{codebox}
\begin{Verbatim}[fontsize=\scriptsize]
Turn 5 [User]:
Perform a detailed analysis of the architecture and design patterns.

Turn 6 [Assistant]:
Let me continue exploring critical components for comprehensive analysis.

Tool Calls:
  - read_file("EduGate_ScholarLink//src//http_handler.c")
  - read_file("EduGate_ScholarLink//src//components//monitoring.c")
  - search_code("route_table", "EduGate_ScholarLink")
  - search_code("pthread", "EduGate_ScholarLink")

Architecture Analysis:
  1. Single-threaded initialization, multi-threaded request handling
  2. No synchronization primitives for shared route_table
  3. Config parser is tightly coupled to initialization
  4. No file watching mechanism for gateway.conf
\end{Verbatim}
\end{codebox}

\newpage
\textbf{Phase 3: Implementation (Turns 7-10)}

\begin{codebox}
\begin{Verbatim}[fontsize=\scriptsize]
Turn 7 [User]:
Based on your analysis, implement improvements to demonstrate understanding.

Turn 8 [Assistant]:
I'll implement a complete hot-reload solution with thread-safe route updates.

Tool Calls:
  - write_file("EduGate_ScholarLink//include//config_reloader.h")
    Content: Declares config_reloader_t with pthread_rwlock, file watcher
  - write_file("EduGate_ScholarLink//src//components//config_reloader.c")
    Content: Implements inotify-based file watching, RCU-style route swapping
  - write_file("EduGate_ScholarLink//src//components//router_v2.c")
    Content: Thread-safe route lookup with read locks, atomic route_table swap

Implementation Strategy:
  • Use pthread_rwlock for read-optimized concurrent access
  • inotify() to watch gateway.conf for IN_CLOSE_WRITE events
  • Parse new config → validate → atomic pointer swap → free old routes
  • Signal handling (SIGHUP) as alternative trigger mechanism

Turn 9 [User]:
Continue implementing improvements. Use write_file to create/modify code.

Turn 10 [Assistant]:
Completing the implementation with integration into main.c and tests.

Tool Calls:
  - write_file("EduGate_ScholarLink//src//main_with_hotreload.c")
    Content: Updated main() to spawn config_reloader thread, register SIGHUP
  - write_file("EduGate_ScholarLink//tests//test_hot_reload.c")
    Content: Unit tests for concurrent route lookups during reload
  - run_tests(language="c", test_path="tests/test_hot_reload.c")
    Result: 8/8 tests passed

Final Architecture:
  • Main thread: HTTP request handling with rwlock-protected route lookups
  • Reloader thread: inotify loop, parse-validate-swap on config changes
  • Zero downtime: existing requests use old routes, new requests use new routes
  • Graceful degradation: invalid config rejected, old routes retained

Session Complete: Architectural refactoring implemented with thread-safe 
hot-reload mechanism, validated through tests.
\end{Verbatim}
\end{codebox}

\newpage
\subsubsection{Example 2: Python Language - GPT-5}

\textbf{Scenario:} Python System Monitoring - Code Comprehension (Expert)  \\
\textbf{Task:} Trace prometheus metric export flow in distributed monitoring system.

\textbf{Phase 1: Initial Exploration}

\begin{codebox}
\begin{Verbatim}[fontsize=\scriptsize]
Turn 1 [User - Initial Prompt]:
Explore the codebase and understand how metrics are collected, aggregated, and 
exported to Prometheus in this distributed system monitoring application.

Turn 2 [Assistant]:
I'll systematically explore the monitoring system architecture to understand the
complete metric flow from collection to Prometheus export.

Tool Calls:
  - list_directory() → Found: main.py, metrics/, exporters/, config/
  - read_file("main.py") → Multi-process worker architecture
  - read_file("metrics/collector.py") → Custom metric collection logic
  - read_file("config/prometheus.yaml") → Export configuration

Key Findings:
  • 4 worker processes collecting system metrics independently
  • Shared memory region for cross-process metric aggregation
  • Custom Prometheus registry with multiprocess mode
  • Export endpoint on port 9090

Turn 3 [User]:
Continue exploring the metric export mechanism and aggregation strategy.
\end{Verbatim}
\end{codebox}

\newpage
\textbf{Phase 2: Deep Analysis}

\begin{codebox}
\begin{Verbatim}[fontsize=\scriptsize]
Turn 4 [Assistant]:
Let me examine the Prometheus integration layer and aggregation mechanism.

Tool Calls:
  - read_file("exporters/prometheus_exporter.py")
  - search_code("exposition", "exporters/")
  - grep_search("REGISTRY")
  - read_file("metrics/aggregator.py")

Technical Analysis:
  • Uses prometheus_client.MultiProcessCollector for process-safe metrics
  • Shared memory mapped files in /tmp/prometheus_multiproc_dir/
  • Custom collectors for: CPU, memory, disk I/O, network stats
  • Atomic counter updates using multiprocessing.Value with ctypes
  • Aggregation strategies: SUM for counters, AVG for gauges, MAX for peaks

Turn 5 [User]:
Document your understanding of the complete metric flow.

Turn 6 [Assistant]:
Creating comprehensive documentation of the metric pipeline.

Tool Calls:
  - write_file("docs/metric_architecture.md")
  - write_file("docs/prometheus_integration.md")

Documentation Created:
  1. Data Flow Diagram: Worker → Collector → Aggregator → Exporter → Prometheus
  2. Multiprocess Safety: Lock-free counters, memory-mapped file coordination
  3. Export Format: OpenMetrics text exposition format
  4. Performance: ~50$\mu$s per metric update, 10K metrics/sec throughput

Session Complete: Comprehensive understanding documented with architecture
diagrams, data flow analysis, and performance characteristics.
\end{Verbatim}
\end{codebox}

\newpage
\subsubsection{Example 3: Rust Language - Gemini-2.5-Pro}

\textbf{Scenario:} Rust Web Dashboard - Bug Investigation (Expert)  \\
\textbf{Task:} Debug WebSocket connection drops under high concurrent load.

\textbf{Phase 1: Problem Investigation}

\begin{codebox}
\begin{Verbatim}[fontsize=\scriptsize]
Turn 1 [User - Initial Prompt]:
Investigate intermittent WebSocket connection drops when 100+ clients connect
simultaneously to the real-time dashboard. Connections drop after 30-60 seconds
under load, but work fine with <50 clients.

Turn 2 [Assistant]:
I'll analyze the WebSocket handling, connection management, and identify the
root cause of connection instability under high load.

Tool Calls:
  - read_file("src/ws_handler.rs") → Connection handler implementation
  - read_file("src/connection_pool.rs") → Pool management logic
  - search_code("tokio::spawn") → Async task spawning patterns
  - grep_search("unbounded") → Channel type usage

Issue Found:
  • Unbounded channels (tokio::sync::mpsc::unbounded_channel) for message passing
  • No backpressure mechanism when clients can't keep up
  • Memory grows linearly with message backlog (observed: 2GB+ under load)
  • Channel full → tokio tasks block → eventually timeout and disconnect

Turn 3 [User]:
Design and implement a fix for the connection stability issue.
\end{Verbatim}
\end{codebox}

\newpage
\textbf{Phase 2: Implementation}

\begin{codebox}
\begin{Verbatim}[fontsize=\scriptsize]
Turn 4 [Assistant]:
I'll implement bounded channels with backpressure and connection health monitoring.

Tool Calls:
  - write_file("src/ws_handler_v2.rs")
    Content: Bounded mpsc channels (capacity: 1000 messages), async send with
    timeout, connection state tracking
  - write_file("src/backpressure.rs")
    Content: Backpressure strategies - drop oldest for real-time data, block for
    critical updates, client speed classification
  - write_file("src/health_monitor.rs")
    Content: Per-connection health tracking, automatic cleanup of slow consumers

Solution Architecture:
  • tokio::sync::mpsc::channel(1000) replaces unbounded channels
  • 3-tier backpressure: fast (no delay), medium (warn client), slow (disconnect)
  • Memory limit per connection: 10MB maximum
  • Health check: ping/pong every 30s, disconnect if 3 consecutive failures

Turn 5 [User]:
Test the fix and validate stability under load.

Turn 6 [Assistant]:
Running comprehensive load tests.

Tool Calls:
  - run_tests(test_path="tests/ws_load_test.rs")
  - run_tests(test_path="tests/ws_integration_test.rs")

Test Results:
  • Load Test (200 clients): 0 drops over 10 minutes, memory stable at 180MB
  • Slow Client Test: Graceful degradation, no impact on other connections
  • Reconnection Test: Clients auto-reconnect, no data loss for critical messages
  • Performance: p50 latency 8ms, p99 latency 45ms (was 2000ms+)

Session Complete: WebSocket stability fixed, production deployment ready.
\end{Verbatim}
\end{codebox}

\newpage
\subsubsection{Example 4: TypeScript Language - Claude-Sonnet-4.5}

\textbf{Scenario:} TypeScript Blockchain NFT - Feature Implementation (Expert) \\ 
\textbf{Task:} Implement batch minting with gas optimization for NFT marketplace.

\textbf{Phase 1: Analysis}

\begin{codebox}
\begin{Verbatim}[fontsize=\scriptsize]
Turn 1 [User - Initial Prompt]:
Implement efficient batch minting functionality that minimizes gas costs when
users mint multiple NFTs in a single transaction. Current implementation requires
separate transactions for each NFT, which is expensive for creators.

Turn 2 [Assistant]:
I'll analyze the current minting contract and identify gas optimization opportunities.

Tool Calls:
  - read_file("contracts/NFTMarketplace.sol") → Current implementation
  - read_file("src/services/mint.service.ts") → Frontend minting service
  - search_code("mintNFT") → All minting call sites
  - search_code("_safeMint") → OpenZeppelin usage

Current State Analysis:
  • Individual _safeMint() calls for each NFT (21K gas per mint)
  • Separate metadata storage writes for each token
  • Redundant ownership transfers and event emissions
  • No bulk discount mechanism

Turn 3 [User]:
Design and implement an optimized batch minting solution.
\end{Verbatim}
\end{codebox}

\newpage
\textbf{Phase 2: Implementation}

\begin{codebox}
\begin{Verbatim}[fontsize=\scriptsize]
Turn 4 [Assistant]:
Creating optimized batch minting with ERC-721 extensions and gas-efficient patterns.

Tool Calls:
  - write_file("contracts/BatchMintable.sol")
    Content: Uses ERC721Enumerable, unchecked increment for token IDs, single
    base URI storage update, loop unrolling for small batches
  - write_file("src/services/batch-mint.service.ts")
    Content: TypeScript SDK with transaction batching, gas estimation, progress
    tracking for large batches
  - write_file("contracts/optimizations/GasOptimizer.sol")
    Content: Helper functions for efficient metadata packing

Gas Optimization Strategies:
  • Single SSTORE for base metadata URI + token range
  • Unchecked arithmetic in loops (Solidity 0.8.0+)
  • Batch event emission (single event with array)
  • Calldata optimization: use arrays instead of individual parameters

Turn 5 [User]:
Test the implementation and measure gas savings.

Turn 6 [Assistant]:
Running gas benchmarks and integration tests.

Tool Calls:
  - run_tests(test_path="test/BatchMinting.test.ts")

Gas Benchmark Results:
  • 1 NFT: 21,000 gas (baseline)
  • 10 NFTs: 73,500 gas (7,350 per NFT, 65% savings)
  • 100 NFTs: 468,000 gas (4,680 per NFT, 78% savings)
  • 1000 NFTs: 4,124,000 gas (4,124 per NFT, 80% savings)

Session Complete: Batch minting deployed to testnet, documentation and frontend
integration completed.
\end{Verbatim}
\end{codebox}

\newpage
\subsubsection{Example 5: Go Language - GPT-5}

\textbf{Scenario:} Go ML NLP - Multi-Session Development (Expert)  \\
\textbf{Task:} Add streaming inference support for transformer-based language models.

\textbf{Phase 1: Analysis}

\begin{codebox}
\begin{Verbatim}[fontsize=\scriptsize]
Turn 1 [User - Initial Prompt]:
Implement streaming token generation for the NLP inference service to enable
real-time text generation with progress updates. Currently only batch inference
is supported with blocking calls.

Turn 2 [Assistant]:
I'll examine the current inference pipeline and gRPC service architecture to
understand how to add streaming support.

Tool Calls:
  - read_file("internal/inference/model.go") → Batch inference implementation
  - read_file("api/proto/nlp_service.proto") → RPC service definitions
  - list_directory("api/proto/") → Available protobuf files
  - search_code("UnaryRPC") → Current RPC patterns

Current Architecture:
  • Batch-only inference with blocking RPC calls
  • No intermediate result streaming
  • Client waits for complete generation (30-60s for long texts)
  • No progress feedback or cancellation support

Turn 3 [User]:
Design the streaming architecture with proper error handling.
\end{Verbatim}
\end{codebox}

\newpage
\textbf{Phase 2: Implementation}

\begin{codebox}
\begin{Verbatim}[fontsize=\scriptsize]
Turn 4 [Assistant]:
I'll implement server-side streaming gRPC with token-by-token generation.

Tool Calls:
  - write_file("api/proto/streaming.proto")
    Content: StreamTokensRequest/Response messages, server streaming RPC definition
  - write_file("internal/inference/streaming_model.go")
    Content: Token generator with goroutine-based async generation
  - write_file("internal/inference/token_stream.go")
    Content: Buffered channel (cap: 100), backpressure handling

Solution Architecture:
  • gRPC server streaming for progressive token delivery
  • Goroutine-based token generator with context cancellation
  • Buffered channels for backpressure (prevents memory overflow)
  • Graceful shutdown on client disconnect

Turn 5 [User]:
Integrate with existing services and add observability.

Turn 6 [Assistant]:
Adding monitoring and completing integration.

Tool Calls:
  - write_file("internal/server/streaming_handler.go")
  - write_file("internal/metrics/streaming_metrics.go")
  - write_file("cmd/server/main.go") [integration with existing server]

Observability:
  • Prometheus metrics: tokens_generated_total, stream_latency_seconds,
    active_streams_gauge, stream_errors_total
  • Structured logging for stream lifecycle events
  • Distributed tracing with OpenTelemetry

run_tests() → All 28 tests passed
  • Unit tests: streaming logic, backpressure, cancellation
  • Integration tests: end-to-end streaming, error scenarios
  • Performance: p50=12ms, p99=45ms per token

Session Complete: Streaming inference deployed to staging, monitoring
dashboards configured, documentation updated.
\end{Verbatim}
\end{codebox}

\newpage
\subsubsection{Example 6: Java Language - Gemini-2.5-Pro}

\textbf{Scenario:} Java ML Training - Cross-File Refactoring (Expert)  \\
\textbf{Task:} Refactor tightly coupled feature pipeline for pluggable transformations.

\textbf{Phase 1: Architecture Analysis}

\begin{codebox}
\begin{Verbatim}[fontsize=\scriptsize]
Turn 1 [User - Initial Prompt]:
Refactor the feature engineering pipeline to support pluggable transformers
without modifying core orchestration code. The current system has 12 hard-coded
transformers that require recompilation for any changes.

Turn 2 [Assistant]:
I'll analyze the current pipeline architecture and identify coupling points.

Tool Calls:
  - read_file("FeaturePipeline.java") → Pipeline orchestration logic
  - read_file("TrainingOrchestrator.java") → Training workflow
  - search_code("FeatureTransformer") → Transformer usage patterns
  - search_code("implements") → Current interface implementations
  - grep_search("new.*Transformer") → Direct instantiation points

Architectural Issues Identified:
  • 12 hard-coded transformers with direct instantiation
  • Tight coupling between pipeline and concrete transformer classes
  • No dependency injection framework
  • Modifications require recompiling orchestration layer
  • No plugin discovery mechanism

Turn 3 [User]:
Design a flexible plugin architecture with backwards compatibility.
\end{Verbatim}
\end{codebox}

\newpage
\textbf{Phase 2: Plugin System Implementation}

\begin{codebox}
\begin{Verbatim}[fontsize=\scriptsize]
Turn 4 [Assistant]:
I'll implement a strategy pattern with Spring dependency injection and plugin
discovery.

Tool Calls:
  - write_file("FeatureTransformerPlugin.java")
    Content: Plugin interface with transform(), validate(), getPriority() methods
  - write_file("TransformerRegistry.java")
    Content: Plugin discovery via Spring @Component scanning, priority-based
    ordering, lifecycle management
  - write_file("PipelineConfig.java")
    Content: Spring configuration for plugin auto-detection

Plugin Architecture:
  • Interface-based plugins (FeatureTransformerPlugin)
  • ServiceLoader for Java SPI discovery (non-Spring deployments)
  • Spring @Component scanning for automatic registration
  • Priority ordering for transformer execution sequence
  • Backwards-compatible adapter for existing transformers

Turn 5 [User]:
Migrate all existing transformers to the new plugin system.

Turn 6 [Assistant]:
Refactoring 12 transformer classes to implement FeatureTransformerPlugin.

Tool Calls:
  - write_file("StandardScalerTransformer.java")
  - write_file("MinMaxNormalizerTransformer.java")
  - write_file("OneHotEncoderTransformer.java")
  - [+9 more transformer files]
  - write_file("example_plugins/CustomEncoderTransformer.java") [example plugin]
  - write_file("docs/PluginDevelopmentGuide.md")

run_tests() → Comprehensive validation:
  • 45 unit tests passed (transformer logic, plugin loading)
  • 12 integration tests passed (end-to-end pipeline with plugins)
  • Backwards compatibility: all existing training jobs work unchanged

Session Complete: Pluggable pipeline architecture deployed to production,
plugin development guide created, 3 community plugins already contributed.
\end{Verbatim}
\end{codebox}

\newpage
\subsubsection{Example 7: JavaScript Language - Claude-Sonnet-4.5}

\textbf{Scenario:} JavaScript System Networking - Integration Testing (Expert) \\ 
\textbf{Task:} Design end-to-end tests for distributed message queue system.

\textbf{Phase 1: Test Design}

\begin{codebox}
\begin{Verbatim}[fontsize=\scriptsize]
Turn 1 [User - Initial Prompt]:
Create comprehensive integration tests for the distributed message queue covering
failure scenarios, network partitions, and consistency guarantees. Current tests
are unit-only.

Turn 2 [Assistant]:
I'll analyze the system architecture and design multi-node integration tests.

Tool Calls:
  - read_file("src/queue/broker.js") → Broker implementation
  - read_file("src/queue/consumer.js") → Consumer logic
  - read_file("src/queue/producer.js") → Producer logic
  - list_directory("tests/") → Existing test structure
  - grep_search("describe") → Current test coverage

Current State:
  • Unit tests only (78% coverage)
  • No multi-node testing infrastructure
  • No failure injection or chaos engineering
  • Untested: network partitions, node failures, consistency edge cases

Turn 3 [User]:
Implement multi-node test framework with Docker and chaos engineering.
\end{Verbatim}
\end{codebox}

\newpage
\textbf{Phase 2: Framework Implementation}

\begin{codebox}
\begin{Verbatim}[fontsize=\scriptsize]
Turn 4 [Assistant]:
Creating Docker-based test harness with Toxiproxy for network failure simulation.

Tool Calls:
  - write_file("tests/integration/cluster-setup.js")
    Content: 5-node cluster orchestration, health checks, cleanup
  - write_file("tests/integration/failure-scenarios.test.js")
    Content: Network partition tests, node crash recovery, split-brain scenarios
  - write_file("docker-compose.test.yml")
    Content: Multi-container setup with Toxiproxy sidecars
  - write_file("tests/integration/chaos-utils.js")
    Content: Network latency injection, packet loss, link cutting

Test Framework:
  • docker-compose for 5-node queue cluster
  • Toxiproxy for controllable network failures
  • Test scenarios: leader election, partition tolerance, message ordering,
    exactly-once delivery, consumer rebalancing

Turn 5 [User]:
Execute the integration test suite and validate distributed guarantees.

Turn 6 [Assistant]:
Running comprehensive distributed system tests.

run_tests() → 45 integration tests passed:
  • Network partitions (3 tests): split-brain prevention, quorum maintenance
  • Node crashes (5 tests): leader failover, data durability
  • Message duplication (8 tests): exactly-once delivery verification
  • Ordering guarantees (7 tests): partition-level ordering, global ordering
  • Consumer rebalancing (10 tests): dynamic group membership
  • Performance under chaos (12 tests): latency, throughput during failures

Session Complete: Integration test suite deployed to CI/CD pipeline,
95% distributed code coverage, chaos tests run nightly.
\end{Verbatim}
\end{codebox}

\newpage
\subsubsection{Example 8: C++ Language - GPT-5}

\textbf{Scenario:} C++ Fintech Trading - Integration Testing (Expert)  \\
\textbf{Task:} Implement latency-sensitive integration tests for HFT order matching.

\textbf{Phase 1: Analysis}

\begin{codebox}
\begin{Verbatim}[fontsize=\scriptsize]
Turn 1 [User - Initial Prompt]:
Create integration tests for the order matching engine with strict latency
requirements (<50$\mu$s p99 for order processing). Current tests don't track
latency percentiles accurately.

Turn 2 [Assistant]:
I'll examine the matching engine and design a low-latency test harness.

Tool Calls:
  - read_file("src/matching_engine.cpp") → Order matching implementation
  - read_file("tests/engine_test.cpp") → Existing test infrastructure
  - search_code("std::chrono") → Current timing mechanisms
  - grep_search("CLOCK_MONOTONIC") → Clock source usage

Current Test Infrastructure Issues:
  • Uses std::chrono (microsecond precision insufficient for <50$\mu$s validation)
  • No percentile tracking (p50/p95/p99/p999)
  • Limited concurrency testing (max 2 threads, production uses 16+)
  • No realistic order flow patterns

Turn 3 [User]:
Design cycle-accurate test harness with HDR histogram for percentile tracking.
\end{Verbatim}
\end{codebox}

\newpage
\textbf{Phase 2: Implementation}

\begin{codebox}
\begin{Verbatim}[fontsize=\scriptsize]
Turn 4 [Assistant]:
Creating TSC-based timing framework with statistical validation.

Tool Calls:
  - write_file("tests/integration/latency_harness.cpp")
    Content: RDTSC wrapper, cycle-to-nanosecond conversion, thread pinning
  - write_file("tests/integration/order_flow_tests.cpp")
    Content: Realistic order patterns, concurrent submissions, SLA validation
  - write_file("tests/integration/tsc_clock.h")
    Content: Inline RDTSC, calibration against CLOCK_MONOTONIC
  - write_file("tests/integration/hdr_histogram.cpp")
    Content: HDR histogram for accurate percentile calculation

Implementation Details:
  • RDTSC (read time-stamp counter) for sub-nanosecond timing accuracy
  • Lock-free SPSC queues for zero-copy test message injection
  • HDR histogram library for percentile calculation (p50/p95/p99/p999)
  • 8-thread concurrent order submission simulating production load
  • Thread pinning to isolate cores and reduce jitter

Turn 5 [User]:
Execute comprehensive latency tests under various load conditions.

Turn 6 [Assistant]:
Running latency benchmarks with different order flow patterns.

run_tests() → Latency Test Results:
  • Low load (100 orders/sec): p50=8.2$\mu$s, p95=18.3$\mu$s, p99=32.1$\mu$s
  • Medium load (1K orders/sec): p50=11.7$\mu$s, p95=28.4$\mu$s, p99=45.2$\mu$s
  • High load (10K orders/sec): p50=12.3$\mu$s, p95=34.7$\mu$s, p99=47.8$\mu$s
  • Spike load (50K burst): p50=15.1$\mu$s, p95=42.3$\mu$s, p999=89.2$\mu$s

All tests passed: p99 latency <50$\mu$s SLA met across all scenarios

Session Complete: Latency validation integrated into CI pipeline with automated
performance regression detection and alerting.
\end{Verbatim}
\end{codebox}

\newpage
\subsubsection{Example 9: C\# Language - Gemini-2.5-Pro}

\textbf{Scenario:} C\# Web CMS - Security Analysis (Expert)  \\
\textbf{Task:} Conduct comprehensive security audit of content management system.

\textbf{Phase 1: Security Audit}

\begin{codebox}
\begin{Verbatim}[fontsize=\scriptsize]
Turn 1 [User - Initial Prompt]:
Perform thorough security analysis of the CMS focusing on authentication,
authorization, input validation, and SQL injection prevention. Target OWASP
Top 10 compliance.

Turn 2 [Assistant]:
I'll systematically review all security-critical components.

Tool Calls:
  - read_file("Controllers/AuthController.cs") → Authentication logic
  - read_file("Services/ContentService.cs") → Data access layer
  - search_code("SqlCommand") → Direct SQL usage
  - grep_search("User.Identity") → Authentication checks
  - search_code("Authorize") → Authorization attributes

Critical Vulnerabilities Identified:
  • SQL injection in search feature (dynamic query string concatenation)
  • Missing CSRF tokens on all POST/PUT/DELETE endpoints
  • Weak password policy (min 6 chars, no complexity requirements)
  • Insecure direct object references in content API (no ownership validation)
  • Sensitive data exposure in error messages (stack traces in production)

Turn 3 [User]:
Prioritize vulnerabilities by CVSS score and implement fixes for critical issues.
\end{Verbatim}
\end{codebox}

\newpage
\textbf{Phase 2: Remediation}

\begin{codebox}
\begin{Verbatim}[fontsize=\scriptsize]
Turn 4 [Assistant]:
Implementing security remediations starting with critical CVSS 9.0+ issues.

Tool Calls:
  - write_file("Services/ContentService.cs")
    Content: Replace raw SQL with Entity Framework parameterized queries
  - write_file("Middleware/CsrfMiddleware.cs")
    Content: Anti-forgery token validation middleware
  - write_file("Models/PasswordPolicy.cs")
    Content: 12+ chars, uppercase, lowercase, numbers, symbols required
  - write_file("Filters/AuthorizeOwnerFilter.cs")
    Content: Resource ownership validation before any CRUD operation

Security Fixes Applied:
  • SQL Injection: Entity Framework with LINQ (no raw SQL)
  • CSRF: ValidateAntiForgeryToken on all state-changing operations
  • Password Policy: ASP.NET Core Identity with custom validator
  • Authorization: Custom filter checking user ID matches resource owner
  • Error Handling: Custom error pages, sanitized messages in production

Turn 5 [User]:
Validate all fixes with automated security scanning and penetration testing.

Turn 6 [Assistant]:
Running OWASP ZAP, Burp Suite scans, and manual penetration tests.

run_tests() → Security Validation Results:
  • SQL injection: 0 vulnerabilities (tested 50+ payloads)
  • CSRF: 0 vulnerabilities (all endpoints protected)
  • Authentication bypass: 0 vulnerabilities
  • Authorization bypass: 0 vulnerabilities
  • OWASP Top 10 compliance: 10/10 checks passed
  • Penetration test: No critical or high severity findings

Session Complete: Security audit report generated with before/after comparisons,
all critical vulnerabilities remediated and verified, quarterly audits scheduled.
\end{Verbatim}
\end{codebox}

\newpage
\subsubsection{Example 10: PHP Language - Claude-Sonnet-4.5}

\textbf{Scenario:} PHP ML NLP - Security Analysis (Expert)  \\
\textbf{Task:} Security hardening of NLP API with user-generated content processing.

\textbf{Phase 1: Threat Assessment}

\begin{codebox}
\begin{Verbatim}[fontsize=\scriptsize]
Turn 1 [User - Initial Prompt]:
Audit and secure the NLP processing API that handles user-submitted text.
Focus on injection attacks, resource exhaustion (DoS), and data leakage.

Turn 2 [Assistant]:
I'll review all API endpoints and identify attack vectors.

Tool Calls:
  - read_file("api/process.php") → Main API endpoint
  - read_file("lib/NLPProcessor.php") → Text processing pipeline
  - search_code("exec") → Command execution calls
  - search_code("shell_exec") → Shell command usage
  - grep_search("passthru") → Other dangerous functions

Critical Security Issues:
  • Command injection: shell_exec() with unsanitized user input in preprocessing
  • No rate limiting: single IP can overwhelm API with requests
  • Information disclosure: model prompts exposed in error messages
  • DoS vector: unbounded input length (accepted 100MB+ text)
  • Secrets in code: API keys hardcoded in NLPProcessor.php

Turn 3 [User]:
Implement defense-in-depth security controls across all layers.
\end{Verbatim}
\end{codebox}

\newpage
\textbf{Phase 2: Security Hardening}

\begin{codebox}
\begin{Verbatim}[fontsize=\scriptsize]
Turn 4 [Assistant]:
Adding comprehensive security controls with layered defense.

Tool Calls:
  - write_file("lib/InputValidator.php")
    Content: Whitelist validation, length limits (10KB), character filtering
  - write_file("middleware/RateLimiter.php")
    Content: Redis-backed rate limiter (100 req/min per IP, 1K/min global)
  - write_file("lib/SecureNLPProcessor.php")
    Content: Replaced shell_exec with safe PHP str functions, input escaping
  - write_file("config/secrets.php")
    Content: Environment variable loading, .env file support
  - write_file("middleware/ErrorHandler.php")
    Content: Sanitized error responses, detailed logging only in dev mode

Security Controls Implemented:
  • Input validation: Whitelist-based (alphanumeric + punctuation only)
  • Length limits: 10KB max input, 100KB max API response
  • Command injection fix: Removed all shell_exec(), used native PHP functions
  • Rate limiting: Token bucket algorithm with Redis (100/min IP, 1000/min global)
  • Secrets management: .env file, never committed to git
  • Error handling: Generic messages in production, detailed logs server-side only

Turn 5 [User]:
Validate security improvements with fuzzing and penetration testing.

Turn 6 [Assistant]:
Running comprehensive security test suite.

run_tests() → Penetration Test Results:
  • Command injection: 0 vulnerabilities (tested 100+ payloads)
  • Rate limiting: Enforced (tested with 500 req/sec attack, properly blocked)
  • Information disclosure: No sensitive data in error responses
  • Input validation: Rejected 15/15 malicious payloads (SQL, XSS, command inj.)
  • DoS resistance: 10KB limit enforced, no memory exhaustion under load

Session Complete: Security hardened API deployed to production with Web Application
Firewall (WAF) rules, intrusion detection, and 24/7 monitoring alerts configured.
\end{Verbatim}
\end{codebox}

\end{document}